\documentclass[reprint,aps,prl,superscriptaddress,floatfix]{revtex4-1}
\usepackage[utf8x]{inputenc}
\usepackage{amsmath}
\usepackage{graphicx}
\usepackage{amssymb}
\usepackage{txfonts}
\usepackage{color}

\newcommand{\dt}{\ensuremath{\partial_t}}

\newcommand{\fn}{\ensuremath{\mathbf n}}
\newcommand{\fnr}{\ensuremath{\mathbf n_{\mathbf r}}}
\newcommand{\fr}{\ensuremath{\mathbf r}}
\newcommand{\fnrx}{\ensuremath{\mathbf n_{\mathbf r + \mathbf e_x}}}
\newcommand{\fnry}{\ensuremath{\mathbf n_{\mathbf r + \mathbf e_y}}}
\newcommand{\vdn}{\ensuremath{ (\mathbf v_s \cdot\nabla)}}
\newcommand{\Beff}{\ensuremath{\mathbf B_{\rm eff}}}
\newcommand{\bpdn}{\ensuremath{b_{\perp 1}}}
\newcommand{\bpndn}{\ensuremath{b_{\perp 2}}}

\newcommand{\fjskt}{\ensuremath{\mathbf j_{\rm Sk}^{\rm (1)}}}
\newcommand{\fjskd}{\ensuremath{\mathbf j_{\rm Sk}^{\rm (2)}}}
\newcommand{\hvdn}{\ensuremath{ (\hat{\mathbf v} \cdot\nabla)\fn}}
\newcommand{\hvpdn}{\ensuremath{ (\hat{\mathbf v}_{\perp} \cdot\nabla)\fn}}

\begin{document}

\title{Skyrmion-Antiskyrmion pair creation by in-plane currents}

\author{Martin Stier}
\affiliation{I. Institut f\"ur Theoretische Physik, Universit\"at Hamburg,
Jungiusstra{\ss}e 9, 20355 Hamburg, Germany}
\author{Wolfgang H\"ausler}
\affiliation{I. Institut f\"ur Theoretische Physik, Universit\"at Hamburg,
Jungiusstra{\ss}e 9, 20355 Hamburg, Germany}
\affiliation{Institut f\"ur Physik, Universit\"at Augsburg, 86135 Augsburg, Germany}
\author{Thore Posske}
\author{Gregor Gurski}
\author{Michael Thorwart}
\affiliation{I. Institut f\"ur Theoretische Physik, Universit\"at Hamburg,
Jungiusstra{\ss}e 9, 20355 Hamburg, Germany}

\begin{abstract}
Magnetic skyrmions can be considered as topologically protected localized vortex-like spin textures. 
Due to their stability, their small size, and the possibility to move them by low electric currents 
they are promising candidates for spintronic devices. Without violating topological protection, it 
is possible to create skyrmion-antiskyrmion pairs, as long as the 
total charge remains unchanged. We derive a skyrmion 
equation of motion which 
reveals how spin-polarized charge currents create skyrmion-antiskyrmion pairs. It 
allows to identify general prerequisites for the pair creation 
process. We corroborate these general principles by numerical 
simulations. On a lattice, where 
topological protection becomes imperfect, the antiskyrmion partner of the pairs is annihilated
and only the skyrmion survives. This eventually changes the total skyrmion number and 
yields a new way of creating and controlling skyrmions.
\end{abstract}


\maketitle
Magnetic skyrmions (Sks) are vortex-like localized magnetization configurations 
 \cite{SKYRME1962556,kiselev2011chiral} which have been predicted
\cite{bogdanov1994thermodynamically} before they were discovered experimentally
\cite{yu2011near,heinze2011spontaneous,hanneken2015electrical,muhlbauer2009skyrmion} in 
magnetic layers with a strong spin-orbit interaction
\cite{bogdanov95,bogdanov1999stability,rossler2006spontaneous}. Despite their potentially small size 
\cite{shibata2013towards,romming2015field}, their thermodynamic
stability is considerably strong
\cite{muhlbauer2009skyrmion,heinze2011spontaneous,hagemeister2015stability}. This is 
a consequence of the particular magnetic configuration which can be characterized by a total
topological charge or Sk number $Q$. 
It can take 
integer values
only and therefore cannot be changed continuously 
\cite{QiQuZhang2006TopologicalQuantizationOfTheSpinHallEffectIn2DParamagneticSemiconductors, 
Hirsch1976DifferentialTopology}. This feature
protects magnetic Sks against typical 
drawbacks of solid state systems such
as disorder or imperfect fabrication 
\cite{bogdanov95,bogdanov1999stability,rosch2013skyrmions}.
Together with the property of easy repositioning by rather tiny in-plane electrical currents
\cite{sampaio2013nucleation,iwasaki2013current,iwasaki2013universal,jonietz2010spin,
yu2012skyrmion}, this makes single Sks attractive candidates
for future racetrack memory devices 
\cite{sampaio2013nucleation,krause2016spintronics,woo2016observation,zhang2015skyrmion,
fert2013skyrmions,nagaosa2013topological}. Creation of Sks has been reported 
in the vicinity of 
notches \cite{iwasaki2013current}, by circular currents \cite{tchoe2012skyrmion}, by 
geometrical constraints \cite{Jiang283} or by sweeping the
external magnetic field \cite{koshibae2016theory}.
Controlled creation and annihilation of individual Sks has been
demonstrated \cite{romming2013writing}. Each of these processes has to overcome topological 
protection and the precise mechanism of each way of Sk creation has to be determined.

In this work, we derive a Sk equation of motion which 
reveals the details of the process how total the topological charge
$Q$ changes by an applied in-plane current. We find this to happen 
in two steps. First, a skyrmion-antiskyrmion (Sk-ASk) pair is created \cite{koshibae2016theory} 
from  small spatial fluctuations of the 
magnetization. Pair creation does not change the total topological 
charge $Q$, since the 
Sk and the ASk have equal topological charge of opposite
sign, respectively. By the externally applied current the Sk and ASk get 
spatially further separated. The Sk equation of motion reveals the relevant terms at work which are 
not captured by the common Thiele approximation \cite{thiele1973steady,muller2015capturing}. 
Finally, the ASk, being no stable solution for a given Zeeman field and a Dzyaloshinsky-Moriya 
interaction,
decays due to Gilbert damping. It is this second step, which is ultimately responsible for changing 
$Q$, crucially relying on dissipation. 
All general findings are confirmed by extended numerical simulations.\\
Recently, Sk-ASk pair creation by in-plane currents in systems without Dzyaloshinsky-Moriya 
interaction has also been reported and investigated numerically \cite{everschor2016skyrmion}.
Additionally, Sk creation by in-plane currents also has been 
observed in experiment \cite{yu2017}.

The two-dimensional magnetization configuration $\mathbf M(x,y,t)$ of
 a single current-driven Sk evolves in time according to the extended
Landau-Lifshitz-Gilbert (LLG) equation 
\cite{tatara2008microscopic,li2004domain,bazaliy1998modification,lakshmanan2011fascinating}
\begin{align}
\dt\fn=&-\fn\times \Beff + \alpha\fn\times\dt\fn\label{eq::llg}\\
& + \vdn\fn - \beta \fn\times\vdn\fn\nonumber
\end{align}
where $\fn = \mathbf M / |\mathbf M|$ is a normalized vector
field. All interactions of the Hamiltonian $H$ describing the
system are contained in the effective field $\Beff = -\partial
H/\partial\fn$. Below, in Eq.~(\ref{latticemodel}), we specify the Hamiltonian 
for a lattice model, but its detailed form is 
not relevant for the following consideration. $\Beff$ already contains 
the gyromagnetic ratio and we set $\hbar=1$. 
Further important parameters are the Gilbert damping constant $\alpha$ and
the non-adiabaticity parameter $\beta$. In this work, we focus specifically
on the impact of spin-polarized electric currents $\mathbf v_s= p
a^3\mathbf j_c /(2e)$ \cite{zhang2009generalization} flowing in the 
magnetic plane with spin polarization $p$
and lattice constant $a$, proportional to a charge current
density $\mathbf j_c$. With the vector field $\fn(x,y,t)$, we then define 
the topological charge density
\begin{equation}\label{eq::qxyt}
q(x,y,t)= \fn\cdot[\hvdn\times\hvpdn]\;,
\end{equation}
and the total topological charge
\begin{equation}
Q=Q(t)=\frac 1 {4\pi}\int {\rm d}^2\fr\ q(x,y,t)\;,\quad Q\in 
\mathbb{Z}\ .\label{eq::Q}
\end{equation}
In fact, this homotopy invariant completely determines the topological 
properties of Sks even though it does not specify, e.g, the 
vorticity of a Sk (ASk) without further 
definitions \cite{
QiQuZhang2006TopologicalQuantizationOfTheSpinHallEffectIn2DParamagneticSemiconductors, 
Hirsch1976DifferentialTopology,heo2016switching}. In our work, however, the magnetic background will 
be fixed in such a way that $ Q > 0$ ($Q < 0$) refers to skyrmions 
(antiskyrmions) \cite{bogdanov1989thermodynamically}. 
For convenience, we take the direction of the spin
current $\mathbf v_s$ as reference direction, $\hat{\mathbf
v} = \mathbf v_s /|\mathbf v_s|$ and $\hat{\mathbf
v}_{\perp}=\hat{\mathbf z} \times \hat{\mathbf v}$. While the topological invariant $Q$ is 
conserved in time at low energies,  the time evolution of $q(x,y,t)$ describes the 
current-induced local motion of Sks. In particular, 
as discussed below, it also describes the generation or annihilation of Sk-ASk
pairs.

To reveal the Sk-ASk pair creation mechanism, we decompose the effective field 
according to 
\begin{equation}
\Beff = b_{\varparallel}\fn + \bpdn\vdn\fn + \bpndn\fn\times\vdn\fn \, .
\label{eq::basis}
\end{equation}
By combining Eqs.\ (\ref{eq::llg}), (\ref{eq::qxyt}) and (\ref{eq::basis}), 
we readily obtain the Sk equation of motion 
\begin{equation}
\dt q = - \nabla\cdot \left(\fjskt+\fjskd\right) \, , 
\label{eq::eom}
\end{equation}
with the Sk currents 
\begin{subequations}\label{eq::jsk}
\begin{align}
\fjskt =& -j_1 q \mathbf v_s\, , \label{eq::jskt}\\
\fjskd =& j_2 \Big\{[\hvdn\cdot\hvpdn]\mathbf v_s\label{eq::jskd1}\\
&\qquad-[\hvdn]^2\mathbf v_{\perp}\Big\}\, , \label{eq::jskd2}
\end{align}
\end{subequations}
 with contributions parallel to the current flow ($\propto\mathbf v_s$) and perpendicular to 
it ($\propto\mathbf v_{\perp}\equiv\hat{\mathbf z}\times\mathbf v_s$). The coefficients read
 \begin{subequations}\label{eq::jsk_coeff}
\begin{align}
j_1 = & [1+\alpha\beta + \alpha \bpdn + 
\bpndn]/(1+\alpha^2)\, , \label{eq::jt}\\
j_2 = & [\alpha-\beta -
 \bpdn + \alpha \bpndn]/(1+\alpha^2)\ .\label{eq::jd}
\end{align}
\end{subequations}
The Sk equation of motion (\ref{eq::eom}) resembles a continuity equation \cite{Garst2016} which 
connects the topological charge density $q$ with the Sk current density. 
We note, however, that conservation of $Q$ in Eq.\ (\ref{eq::Q})
in the present case is not a consequence of Noether's theorem, albeit conserved
quantities may still exist for Eq.\  (\ref{eq::eom})
\cite{FREDERICO2007834} under continuous variation of $\fn$
\cite{nagaosa2013topological}.

The physical meaning of $\fjskt$ and $\fjskd$ becomes apparent when we consider
Sks in the steady state where $\dt \fn = 0$ and thus $\dt q = 0$. For not too large current 
densities, no major structural changes of the magnetization occur and 
$\Beff$ remains parallel to $\fn$. Then, the perpendicular 
components $\bpdn=\bpndn$ vanish and the coefficients $j_1$ and $j_2$ in Eq.\ 
(\ref{eq::jsk_coeff}) 
are constant. A case of special importance occurs when $\alpha=\beta$, which 
implies that 
 $j_1=1$, $j_2=0$. Then, $\fjskd=0$, such that the
undistorted topological charge density $q$ moves with the velocity $-\mathbf
v_s$, according to Eq.\ (\ref{eq::jskt}). This motivates us to call $\fjskt$ a Sk current density.
When $\alpha\neq\beta$ (but still assuming $\bpdn=\bpndn=0$), 
$j_2$ becomes nonzero. Then, we may rewrite
Eqs.~(\ref{eq::jskd1},\ref{eq::jskd2}) in the form $\fjskd =
-j_2q(\eta_{\varparallel}\mathbf v_s + \eta_{\perp}\mathbf
v_{\perp})$, with the coefficients $\eta_{\varparallel} = \hvdn\cdot\hvpdn/q$ and
$\eta_{\perp} = -[\hvdn]^2/q$.
We can simplify $\eta_{\varparallel} =\cot\gamma$, where 
$\gamma$ is the angle between $\hvdn$ and $\hvpdn$.
This term $\propto\eta_{\varparallel}$ only
adds to the contribution of $\fjskt$ (though with a
dependence on the shape of the vector field $\fn$) to drive positive 
and
negative topological charge density along $\pm\mathbf v_s$ depending on the explicit angle 
$\gamma$.\\
Crucial for the following is the term $\propto\eta_{\perp}$. First, 
it points perpendicularly to the externally applied spin
current and, second, it drives negative and positive topological charge densities
in opposite directions, as it changes sign under the inversion $q\to -q$. This is 
essentially the Sk Hall 
effect \cite{nagaosa2013topological,
QiQuZhang2006TopologicalQuantizationOfTheSpinHallEffectIn2DParamagneticSemiconductors,stone1996,
Nakatani2008,jiang2016direct,litzius2016skyrmion}, but for arbitrary topological charge density.
We therefore identify the contribution~(\ref{eq::jskd2}) as being responsible for 
separating negative from positive topological charge density resulting in a common Sk-ASk 
pair. The separation takes place perpendicularly to the external current. 
Actually, this process can be expected to be a common scenario in real materials for 
sufficiently strong applied charge current densities. The only further prerequisites are $\alpha 
-\beta \neq 0$ and small spatial fluctuations of the Sk density $q(x,y,t)$, which also imply  
finite gradients $\hvdn$ and $\hvpdn$ and thus a finite $\fjskd$. A  
finite gradient $\hvpdn$ is, strictly speaking, not necessary for a non-vanishing current $\fjskd$
[cf. Eq. (\ref{eq::jskd2})]. Nevertheless, it is important for a finite divergence 
$\nabla\cdot\fjskd\neq 
0$. Only in this case, the skyrmion current cannot be gauged away and is physically relevant. 
Then, regions of opposite signs appear quite naturally in the 
topologically trivial state $Q=0$, as regions of finite $q$, which we have postulated, have to 
cancel 
each other to sum up to zero. Ultimately, a Sk-ASk pair is formed out 
of these fluctuations. We note in passing that the
detailed motion of ASks is typically more complicated than that of Sks, since 
commonly, an 
isolated ASk is not a stationary solution and thus, already for
$\mathbf v_s=0$,  $\Beff$ is clearly 
not parallel to $\fn$, which implies that $\ensuremath{b_{\perp}}\ne 0$.

\begin{figure*}[t]

\flushright
\includegraphics[width=.9\linewidth]{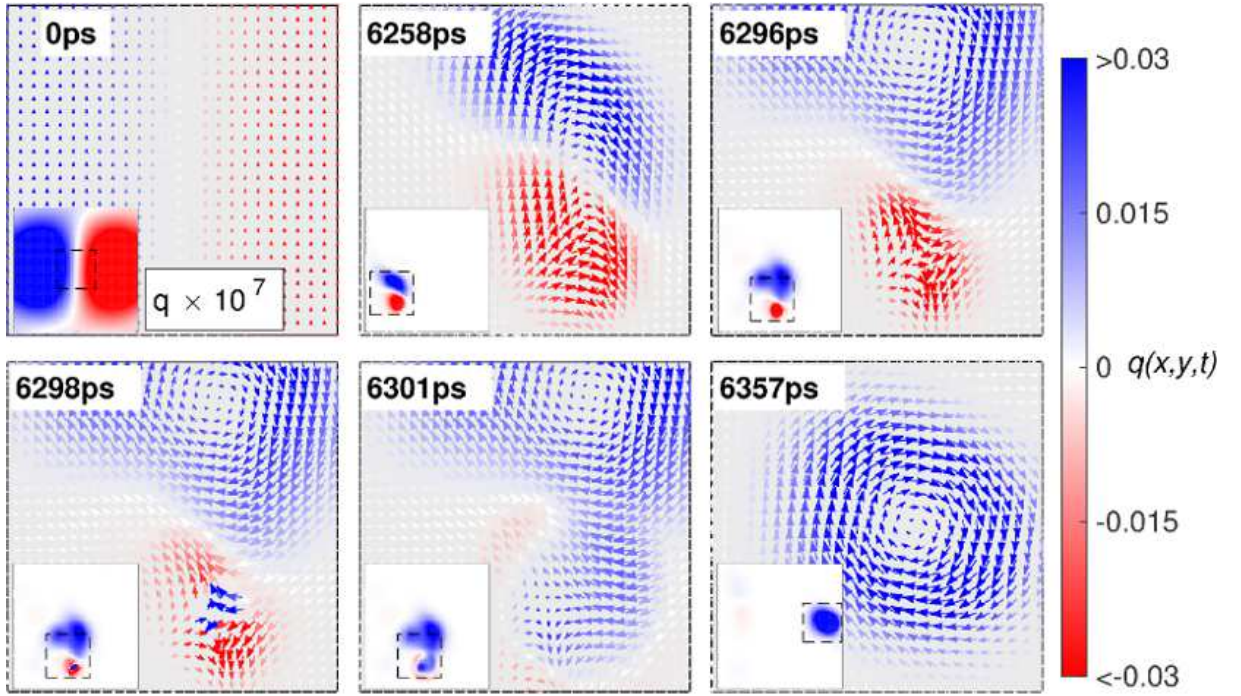}
\caption{\label{fig::snapshots}(Color online). Snapshots of
the topological charge density $q$ (insets) in the relevant $80\times80$ section of the total 
lattice 
and the magnetic texture (arrows) in a magnified section (marked by the dashed rectangles within 
the insets) at the times as indicated. 
The initial topological charge adopted by the applied inhomogeneous 
magnetic field is tiny (see $q$ for $t=0$ which is multiplied by $10^7$ for 
clarity). From these fluctuations the Sk-ASk pair is created 
by a current in x-direction by separating positive and negative topological charge density 
perpendicular to the current direction (here, in the $y$-direction).
The ASk is eventually destroyed around $t\approx 6300\rm ps$ and
only the Sk survives. Parameters are chosen as $I_c=7.7\times10^{11}\rm A/m^2$, $\alpha = 0.25$ and 
$\beta = 0$. Color code refers to topological charge density $q$ always.}
\end{figure*}

In the following, we illustrate these general principles for 
a concrete model realized by the Hamiltonian \cite{iwasaki2013universal}
\begin{align}\label{latticemodel}
H = & -J \sum_{\fr}\fnr\cdot\left(\fnrx+\fnry\right)-\sum_{\fr}\mathbf B_{\mathbf r}\cdot\fnr\\
&-D\sum_{\fr}\left[\left(\fnr\times\fnrx\right)\cdot\mathbf e_x + 
\left(\fnr\times\fnry\right)\cdot\mathbf e_y\right]\nonumber
\end{align}
defined on lattice sites $\fr$ (unit lattice constant) in two
dimensions. It supports Sks in a certain parameter regime of the phase diagram. $J$ is the 
exchange
interaction and $D$ the Dzyaloshinsky-Moriya interaction (DMI)
strength. We use the values $J = 1\rm meV$, $D/J = 0.18$ reported for MnSi
\cite{iwasaki2013universal}. Here, we only discuss a bulk DMI
which stabilizes Bloch Sks. Yet, we have also verified 
our findings for systems with an interfacial DMI which stabilizes 
N\'eel Sks
\cite{thiaville2012dynamics,benitez2015magnetic}. No qualitative 
modifications occur and our findings apply to both kinds of Sks.
In the numerical simulations,
we use a $L_x\times L_y=160\times 160$ square lattice with periodic
boundary conditions. For convenience, we translate $v_s = p a^3 I_c/2e$ into a 
charge current density $I_c$ by assuming full polarization $p=1$ and a lattice constant $a=0.5\rm 
nm$. Depending on
the magnitude of the externally controlled Zeeman field $\mathbf
B$, either a helical phase, a Sk lattice or the
ferromagnetic (field polarized) phase is the ground state
\cite{iwasaki2013universal,koshibae2016theory}. A field
${\mathbf B}=(0,0,B_z)=-0.03\,J\,\hat{\mathbf z}\:$ is in fact
strong enough to align all magnetic moments, $\fn(x,y,t)\equiv
-\hat{\mathbf z}$. Then, $q(x,y,t)$ remains zero everywhere and, according to
Eqs.~(\ref{eq::eom}) and (\ref{eq::jsk}), for all times, since
$\ensuremath{\mathbf j_{\rm Sk}^{\rm (1,2)}=0}$, even at
non-zero applied current densities.

To realize at least a small initial non-zero topological charge density $q$, we add a tiny
 modulation to the magnetic field pointing in the $y$-direction, i.e., $B_y=b_0
[\sin(2\pi x /L_x)+\sin(2\pi y/L_y)]$ and $b_0=B_z/100$. As a matter of fact, the precise 
form of the initial inhomogeneous magnetization configuration is of minor importance.
The time evolution of the system is calculated by solving the extended LLG Eq.\ (\ref{eq::llg}) by  
standard advanced numerical 
methods. 

Starting from the
fully field polarized state $\fn(x,y) \equiv -\hat{\mathbf z}$, we first let
the system accommodate to the additional $B_y$ field, 
at zero external current. After this initial equilibration, we
switch on the current at $t=0$ and calculate $q(x,y,t)$ at every
time step. A movie of this evolution is available in the SM \cite{SM} while a selection of 
snapshots of $q$ is shown in Fig.~\ref{fig::snapshots}. Initially, 
the very small amplitude $b_0$ of $B_y$ generates a tiny seed topological charge density of both 
positive and negative sign with an overall $Q=0$. Gradually, under the influence of
the external current, Sk-ASk pairs begin to form with growing
magnitudes of $q$. Consistent with our theoretical prediction,
the Sk and ASk centers separate in the $y$-direction, perpendicular
to the external current flow. After its full development, since it is unstable,
the ASk disappears on a time scale $\propto 1/\alpha$. Thereby its
diameter shrinks relatively quickly, eventually below the lattice constant.
At this moment, $Q(t)$ abruptly changes by one.\\
As the evolution of Sk-ASk pairs is interfered by 
the relatively short life time of the ASk we further illustrate the details of this process by 
an additional movie \cite{SM} where we set the DMI to zero. Then, neither the Sk nor the ASk is 
energetically preferred and the full Sk-ASk pair evolves in time as recently reported in Ref. 
\cite{everschor2016skyrmion}.

\begin{figure}[t!]
\includegraphics[width=\linewidth,angle=0]{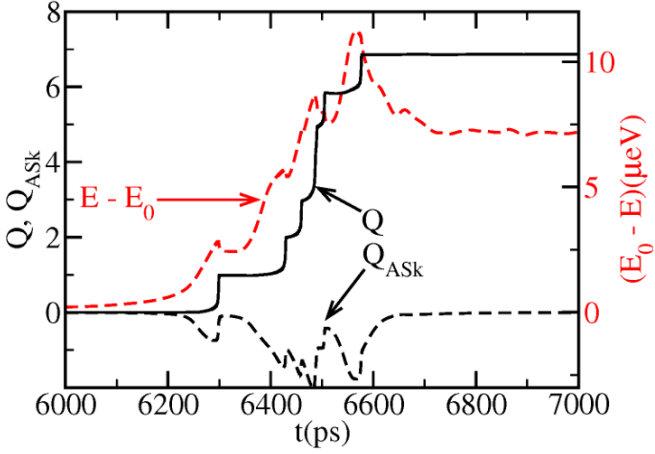}
\caption{\label{fig::Q_vs_t}(Color online). Black solid line: Time-dependence of the total
topological charge $Q(t)$. 
Black dashed line: Time-dependence of the total negative charge 
defined as $Q_{\rm ASk}(t) =\frac 1 {4\pi}
\int_{q<0}{\rm d}x{\rm d}y\:q(x,y,t)$ stemming from negative
$q$ only. Note that $Q_{\rm ASk}$ does not need to be
integer and that the restriction to lattice points
imposes some small, unimportant ambiguity on the precise determination of
$q(x,y,t)$. 
Red dashed line: Time-dependence of the energy in reference to the
initial energy, $E-E_0\equiv E(t)-E(t=0)$, per lattice site. 
Before every $Q$-jump, $Q_{\rm ASk}$ gradually decreases,
accompanied by an increase of the 
energy which eventually is taken from the external
current. Parameters as in Fig. \ref{fig::snapshots}.
}
\end{figure}

The scenario of Sk creation is demonstrated further in 
Fig.~\ref{fig::Q_vs_t}, 
where we show the time-dependence of $Q(t)$. Over large time spans, 
the total topological charge takes an integer values,
while occasionally $Q(t)$ jumps to the next integer within a short
transition time. These transitions are
accompanied by sudden rises of the total negative topological charge 
$Q_{\rm ASk}(t) =\frac 1 {4\pi}
\int_{q<0}{\rm d}x{\rm d}y\:q(x,y,t)$, a
quantity that we define by integrating over negative Sk-density
only. During the times when $Q(t)$ stays integer, $Q_{\rm ASk}(t)$
may decrease gradually with time. This indicates the gradual
creation of Sk-ASk pairs, their growth and their spatial
separation, before the finally isolated, but unstable ASk 
 annihilates during a time much shorter than the
duration of its creation, as described above. 
This initial gradual development of the first Sk-ASk pair due to a weak
spatial inhomogeneity of the Zeeman field is clearly seen
in Fig.~\ref{fig::Q_vs_t}. On the other hand, as soon as a finite number
of Sks exist (after 6300 ps in Fig.~\ref{fig::Q_vs_t}),
their intrinsic inhomogeneous magnetization
suffices to facilitate further creation of Sk-ASk pairs in
their surroundings, even at a homogeneous Zeeman-field
as we have convinced ourselves independently.

Since the system starts very close to the ferromagnetic ground state, the Sk creation costs 
energy. This 
energy is pumped into the system by the charge current.
Figure \ref{fig::Q_vs_t} confirms the connection between the increase of the energy and 
of negative Sk-density.

\begin{figure*}[t]
\begin{minipage}{.45\linewidth}
\includegraphics[width=\linewidth]{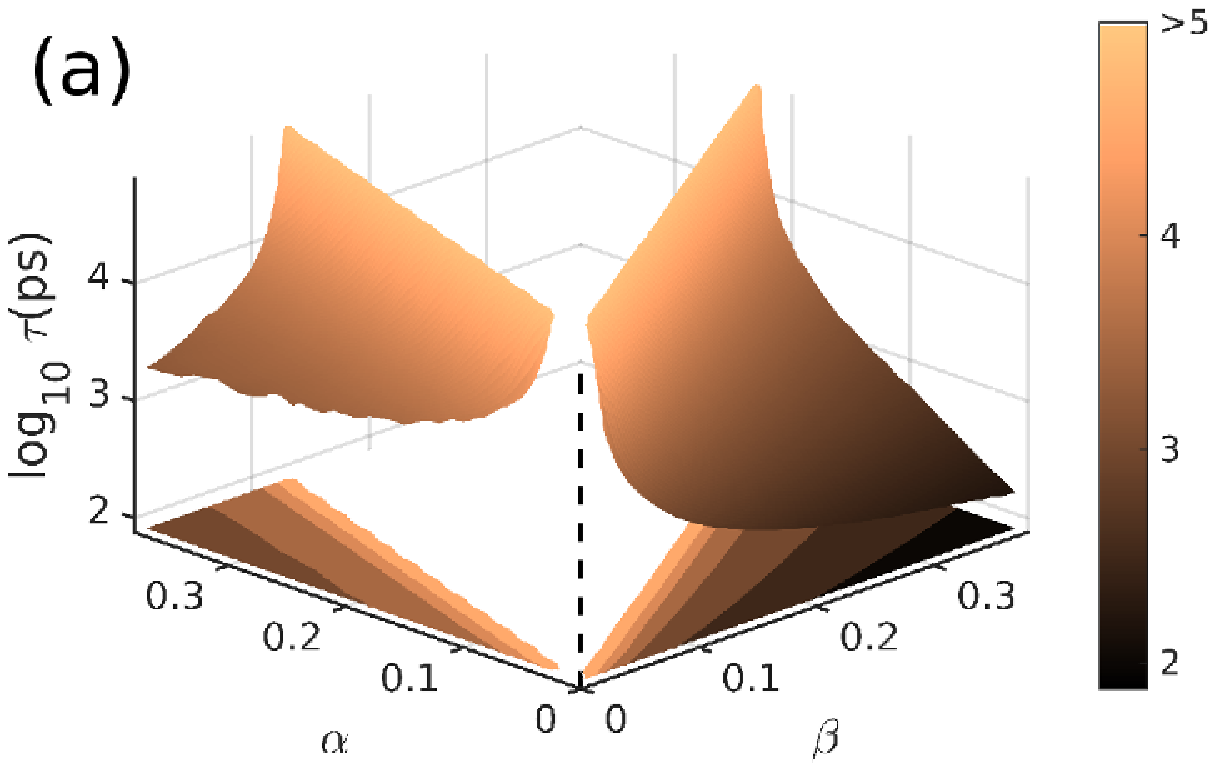}
\end{minipage}
\begin{minipage}{.08\linewidth}
 \
\end{minipage}
\begin{minipage}{.45\linewidth}
\includegraphics[width=\linewidth]{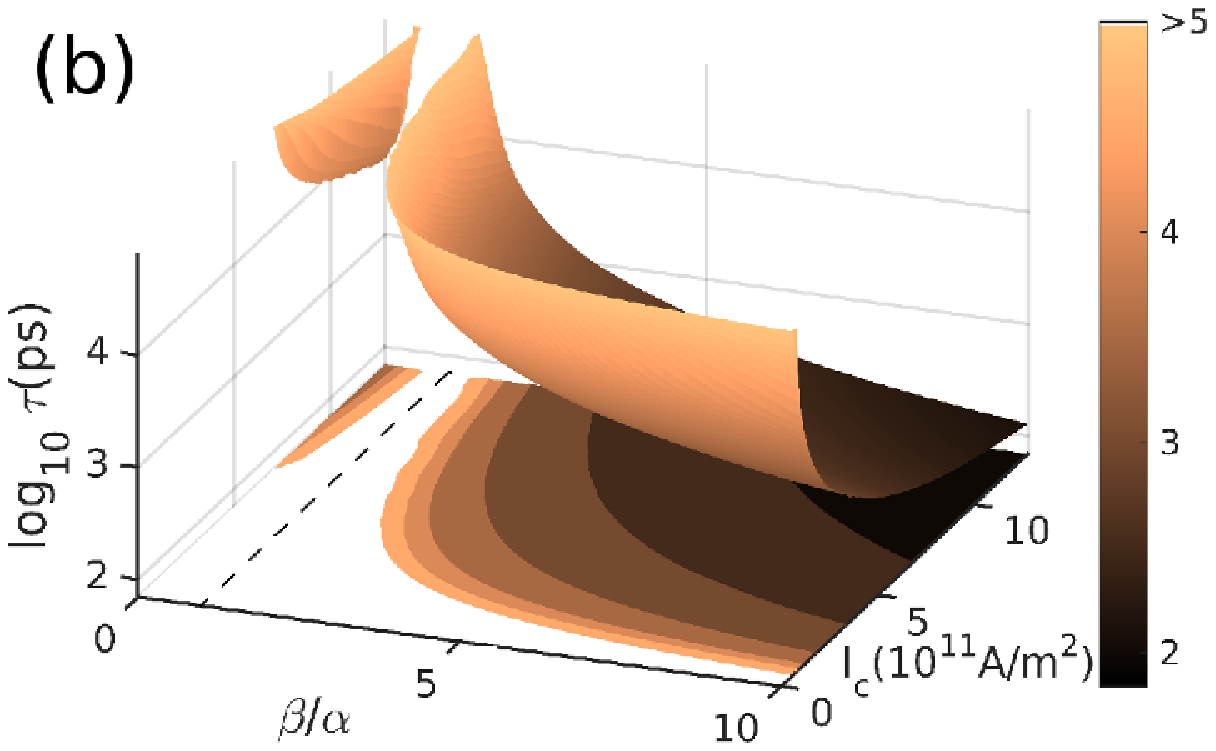}
\end{minipage}
 \caption{\label{fig::tau}(Color online). Decadic logarithm of the Sk creation 
time $\tau$ in dependence of (a) the Gilbert damping constant $\alpha$ and the 
non-adiabaticity parameter $\beta$ for $I_c=10^{12}\rm A/m^2$, and, (b) the ratio $\beta/\alpha$ 
and the charge current density $I_c$ for $\alpha=0.05$. Finite creation times are never achieved at 
$\beta = \alpha$ (dashed lines).  }
\end{figure*}

The duration for Sk creation can be quantified by the time $\tau$ which we define as the time span 
from the 
onset of the current flow till the creation of the first Sk. This 
creation time is 
a combination of the time $\tau_{\rm pair}$ needed to 
form a sufficiently large Sk-ASk pair and the annihilation time $\tau_{\rm ASk}$ 
of the ASk. Since both processes happen at least partially simultaneously, the resulting $\tau$ is 
not a direct sum of both. Still, $\tau_{\rm ASk}\ll\tau_{\rm pair}$
such that we can safely take $\tau \approx \tau_{\rm pair}$. 
Since we attribute the creation of Sk-ASk pairs to the existence of a finite $\fjskd$, we expect 
Sks to be created faster when the magnitude of $\fjskd$ is larger. 
From Eqs.\ (\ref{eq::jskd1}, \ref{eq::jskd2}, \ref{eq::jd}) we find  $|\fjskd| \propto 
(\alpha-\beta) I_c$ in the limit of vanishing 
$\bpdn$ and $\bpndn$. In Fig.\ \ref{fig::tau}, this relation between $\tau$ and $\fjskd$ is 
confirmed 
by the numerical results. Indeed, $\tau$ depends on $|\alpha-\beta|$ and $I_c$. In 
particular, 
no Sks can be created when $\alpha=\beta$ which implies that the dissipative current is 
essential for the charge current-induced Sk creation. Still, finite creation 
times appear in an experimentally relevant parameter regime. Finally, we note that 
even though we have chosen a particular seed magnetic field $B_{y,\mathbf r}$ to create 
topological charge density fluctuations, their precise origin is not important. In 
fact, only an inhomogeneous $q(x,y)$, besides $\alpha\neq\beta$ and $I_c\neq 0$, is 
necessary for $\fjskd$ to become non-vanishing. Thus, a 
multitude of ways
are eligible to create such
fluctuations, for example by local fields, material
modification, or by temperature.
On the other hand, a change of $Q$ will often be undesirable in distinct set-ups. Then,
$\fjskd$-contributions to Eqs.~(\ref{eq::jsk}) should be
suppressed by a proper choice of the material with a small 
$|\alpha-\beta|$, or by avoiding magnetization
fluctuations, apart from simply working in the low current
regime.


In this work, we have established the skyrmion equation of motion by combining the 
general definition of the skyrmion density and the extended Landau-Lifshitz-Gilbert equation.
We here define skyrmion current densities that conserve
the total topological charge of a sample. In the presence of an in-plane spin current, we 
identify terms that give rise to
simple movement of skyrmions against the externally applied
charge flow. Other contributions to skyrmion current densities that we identify explicitly drive 
the
separation of positive skyrmion density from negative antiskyrmion
density perpendicularly to the charge current flow. These latter
contributions eventually cause the creation of skyrmion-antiskyrmion
pairs, already out of very small magnetic 
inhomogeneities. The theoretical predictions are corroborated by numerical simulations and applied
to systems with bulk and interfacial DMI.

\begin{acknowledgments}
We acknowledge support from the DFG SFB 668 (project B16).
\end{acknowledgments}

\appendix

 \section{Detailed spin structure of skyrmionic objects}
 
\begin{figure*}[h!]
\centering
\includegraphics[width=.8\linewidth]{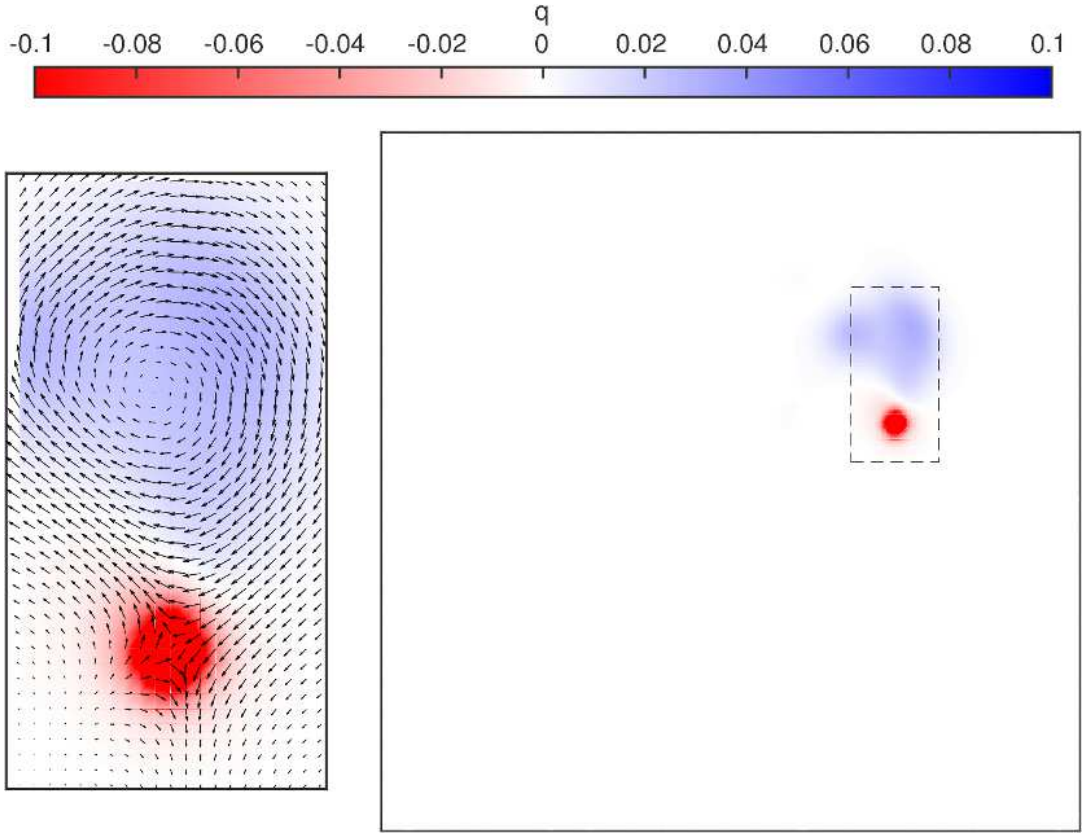}
\caption{\label{fig::bDMI}Skyrmion density $q(x,y,t)$ at $t=6296$ps. The marked area is magnified 
on the left-hand side where the arrows also show the magnetization $n(x,y,t)$ in the $xy$ plane. 
The bulk 
Dzyaloshinskii-Moriya interaction $H_{\rm bDMI} = 
-D\sum_{\fr}\left[\left(\fnr\times\fnrx\right)\cdot\mathbf e_x + 
\left(\fnr\times\fnry\right)\cdot\mathbf e_y\right]$ stabilizes Bloch-like skyrmions.}
\end{figure*}

\begin{figure*}[tb]
\centering
\includegraphics[width=.8\linewidth]{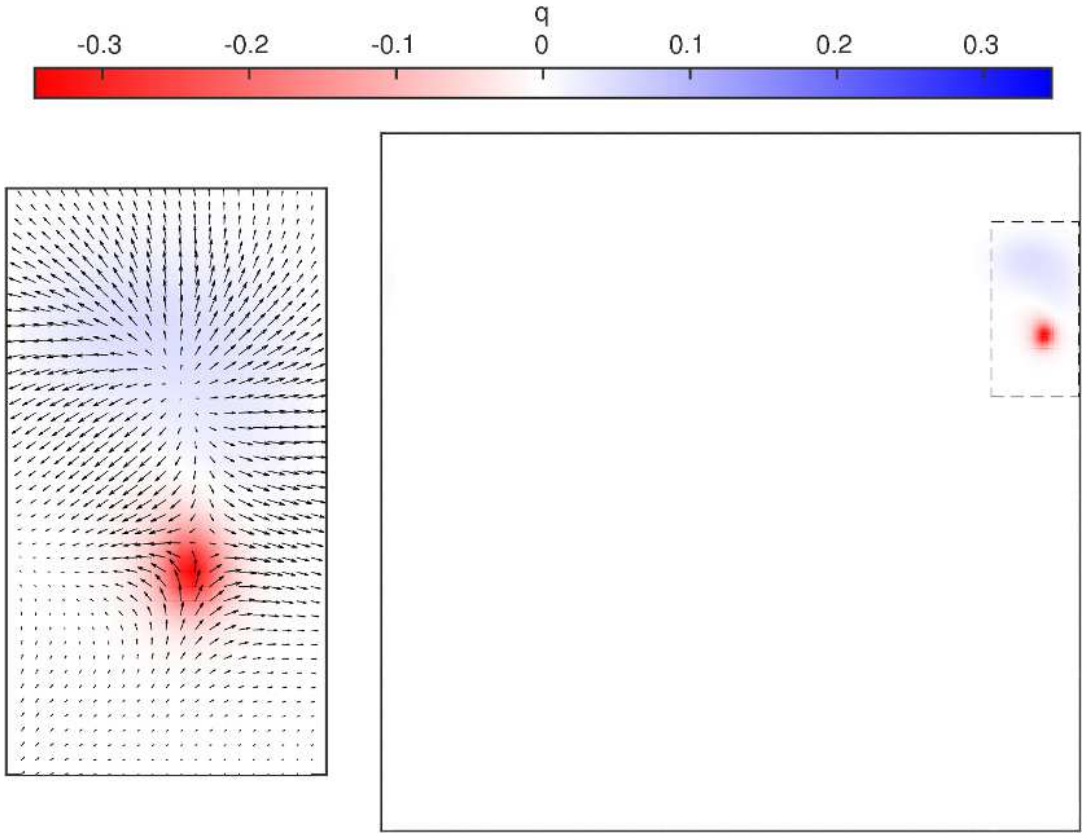}
\caption{\label{fig::iDMI}Skyrmion density $q(x,y,t)$ at $t=4933$ps. The marked area is magnified 
on the 
left-hand side where the arrows also show the magnetization $n(x,y,t)$ in the $xy$ plane. In 
contrast to 
Fig.~\ref{fig::bDMI} an interfacial Dzyaloshinskii-Moriya interaction 
$H_{\rm iDMI} = -D 
\sum_{\fr}\left[(\hat{\mathbf z}\times\hat{\mathbf 
x})\cdot(\fnr\times\fnrx)+(\hat{\mathbf y}\times\hat{\mathbf 
z})\cdot(\fnr\times\fnry)\right]$ has been used, 
which stabilizes Ne\'el-like skyrmions.}

\end{figure*}

\begin{figure*}[tb]
\flushleft (a)\\[-.3cm]
\centering
\includegraphics[width=.75\linewidth]{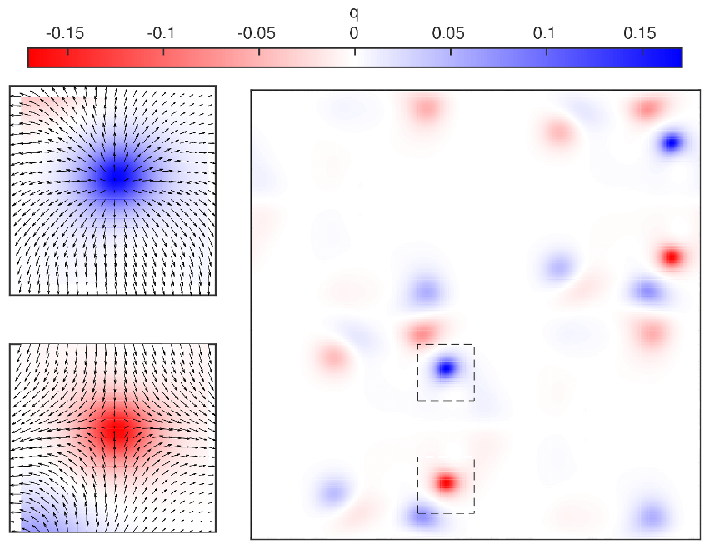}
\flushleft (b)\\[-.3cm]
\centering
\includegraphics[width=.75\linewidth]{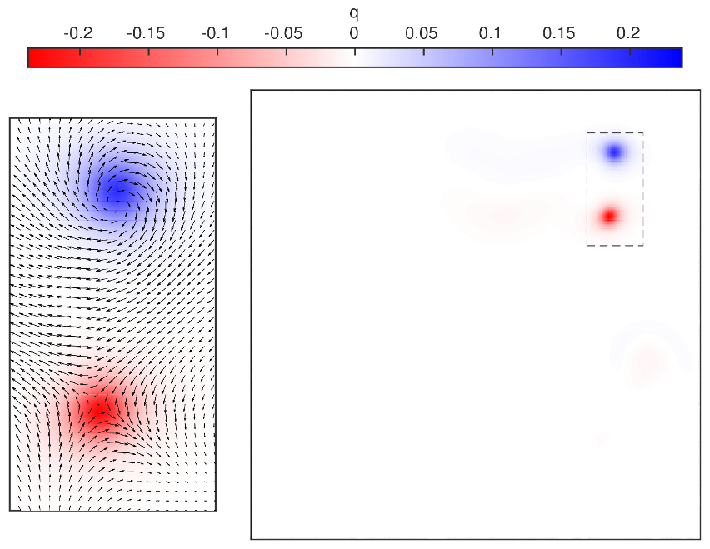}
\caption{\label{fig::zDMI}Skyrmion density $q(x,y,t)$ at (a) $t=10772$ps and (b) $t=14392$ps. The 
marked areas are 
magnified on the left-hand side where the arrows also show the magnetization $n(x,y,t)$ in the 
$xy$ 
plane. 
The Dzyaloshinskii-Moriya interaction is set to zero. This has two consequences: (i) neither the 
skyrmion nor the antiskyrmion is preferred and both types are symmetrically created and (ii) 
skyrmions can be of 
Ne\'el or Bloch shape. }
\end{figure*}

In the main body of the paper, we focus on the dynamics of the skyrmion density as the important 
quantity to describe skyrmions or antiskyrmions, respectively. 
Still, these objects have an internal magnetic structure given by the magnetization $\fn(x,y,t)$. 
In 
particular, different kinds of Dzyaloshinskii-Moriya interactions (DMI) may stabilize different 
types of skyrmions. In this Appendix, we show exemplary magnetic structures for
\begin{itemize}
 \item bulk DMI given by the Hamiltonian  $H_{\rm bDMI} = 
-D\sum_{\fr}\left[\left(\fnr\times\fnrx\right)\cdot\mathbf e_x + 
\left(\fnr\times\fnry\right)\cdot\mathbf e_y\right]$ which stabilizes Bloch-like skyrmions (cf.\ 
Fig.~\ref{fig::bDMI} and movie ``\texttt{SK\_density\_vs\_time\_bulkDMI.avi}''),
 \item interfacial DMI given by the Hamiltonian  $H_{\rm iDMI} = -D 
\sum_{\fr}\left[(\hat{\mathbf z}\times\hat{\mathbf 
x})\cdot(\fnr\times\fnrx)+(\hat{\mathbf y}\times\hat{\mathbf 
z})\cdot(\fnr\times\fnry)\right]$ which stabilizes 
Ne\'el-like skyrmions (cf. Fig.~\ref{fig::iDMI} and movie 
``\texttt{SK\_density\_vs\_time\_interfacialDMI.avi}''), and, 
 \item no DMI with means stabilization neither of Bloch-like nor of Ne\'el-like skyrmions (cf. 
Fig.~\ref{fig::zDMI} and movie ``\texttt{SK\_density\_vs\_time\_zeroDMI.avi}''). 
\end{itemize}
All figures are snapshots taken from movies which we also provide as Supplemental Material online 
in the ``other formats'' option on the article's arXiv page. 
Even though the explicit magnetic structures differ for the according DMI, no qualitative changes 
for the pair creation process, as described in the main article, were observed.\\
We have used the same parameter set as before. The 
calculations were performed on a $L_x\times L_y=160\times 160$ square lattice with periodic
boundary conditions, external magnetic field ${\mathbf B}=(0,0,B_z)=-0.03\,J\,\hat{\mathbf z}\:$, 
spin 
velocity $v_s=-300$m/s, Gilbert damping $\alpha=0.25$ and non-adiabaticity $\beta= 0$. To create 
initial fluctuations of the skyrmion density, a tiny
 modulation to the magnetic field pointing in the $y$-direction, i.e., $B_y=b_0
[\sin(2\pi x /L_x)+\sin(2\pi y/L_y)]$ and $b_0=B_z/100$ has been added to the external field.\\

\begin{thebibliography}{50}%
\makeatletter
\providecommand \@ifxundefined [1]{%
 \@ifx{#1\undefined}
}%
\providecommand \@ifnum [1]{%
 \ifnum #1\expandafter \@firstoftwo
 \else \expandafter \@secondoftwo
 \fi
}%
\providecommand \@ifx [1]{%
 \ifx #1\expandafter \@firstoftwo
 \else \expandafter \@secondoftwo
 \fi
}%
\providecommand \natexlab [1]{#1}%
\providecommand \enquote  [1]{``#1''}%
\providecommand \bibnamefont  [1]{#1}%
\providecommand \bibfnamefont [1]{#1}%
\providecommand \citenamefont [1]{#1}%
\providecommand \href@noop [0]{\@secondoftwo}%
\providecommand \href [0]{\begingroup \@sanitize@url \@href}%
\providecommand \@href[1]{\@@startlink{#1}\@@href}%
\providecommand \@@href[1]{\endgroup#1\@@endlink}%
\providecommand \@sanitize@url [0]{\catcode `\\12\catcode `\$12\catcode
  `\&12\catcode `\#12\catcode `\^12\catcode `\_12\catcode `\%12\relax}%
\providecommand \@@startlink[1]{}%
\providecommand \@@endlink[0]{}%
\providecommand \url  [0]{\begingroup\@sanitize@url \@url }%
\providecommand \@url [1]{\endgroup\@href {#1}{\urlprefix }}%
\providecommand \urlprefix  [0]{URL }%
\providecommand \Eprint [0]{\href }%
\providecommand \doibase [0]{http://dx.doi.org/}%
\providecommand \selectlanguage [0]{\@gobble}%
\providecommand \bibinfo  [0]{\@secondoftwo}%
\providecommand \bibfield  [0]{\@secondoftwo}%
\providecommand \translation [1]{[#1]}%
\providecommand \BibitemOpen [0]{}%
\providecommand \bibitemStop [0]{}%
\providecommand \bibitemNoStop [0]{.\EOS\space}%
\providecommand \EOS [0]{\spacefactor3000\relax}%
\providecommand \BibitemShut  [1]{\csname bibitem#1\endcsname}%
\let\auto@bib@innerbib\@empty
\bibitem [{\citenamefont {Skyrme}(1962)}]{SKYRME1962556}%
  \BibitemOpen
  \bibfield  {author} {\bibinfo {author} {\bibfnamefont {T.}~\bibnamefont
  {Skyrme}},\ }\href {\doibase http://dx.doi.org/10.1016/0029-5582(62)90775-7}
  {\bibfield  {journal} {\bibinfo  {journal} {Nucl. Phys.}\ }\textbf {\bibinfo
  {volume} {31}},\ \bibinfo {pages} {556 } (\bibinfo {year}
  {1962})}\BibitemShut {NoStop}%
\bibitem [{\citenamefont {Kiselev}\ \emph {et~al.}(2011)\citenamefont
  {Kiselev}, \citenamefont {Bogdanov}, \citenamefont {Sch{\"a}fer},\ and\
  \citenamefont {R{\"o}{\ss}ler}}]{kiselev2011chiral}%
  \BibitemOpen
  \bibfield  {author} {\bibinfo {author} {\bibfnamefont {N.}~\bibnamefont
  {Kiselev}}, \bibinfo {author} {\bibfnamefont {A.}~\bibnamefont {Bogdanov}},
  \bibinfo {author} {\bibfnamefont {R.}~\bibnamefont {Sch{\"a}fer}}, \ and\
  \bibinfo {author} {\bibfnamefont {U.}~\bibnamefont {R{\"o}{\ss}ler}},\
  }\href@noop {} {\bibfield  {journal} {\bibinfo  {journal} {J. Phys. D Appl.
  Phys.}\ }\textbf {\bibinfo {volume} {44}},\ \bibinfo {pages} {392001}
  (\bibinfo {year} {2011})}\BibitemShut {NoStop}%
\bibitem [{\citenamefont {Bogdanov}\ and\ \citenamefont
  {Hubert}(1994)}]{bogdanov1994thermodynamically}%
  \BibitemOpen
  \bibfield  {author} {\bibinfo {author} {\bibfnamefont {A.}~\bibnamefont
  {Bogdanov}}\ and\ \bibinfo {author} {\bibfnamefont {A.}~\bibnamefont
  {Hubert}},\ }\href@noop {} {\bibfield  {journal} {\bibinfo  {journal} {J.
  Mag. Mag. Mat.}\ }\textbf {\bibinfo {volume} {138}},\ \bibinfo {pages} {255}
  (\bibinfo {year} {1994})}\BibitemShut {NoStop}%
\bibitem [{\citenamefont {Yu}\ \emph {et~al.}(2011)\citenamefont {Yu},
  \citenamefont {Kanazawa}, \citenamefont {Onose}, \citenamefont {Kimoto},
  \citenamefont {Zhang}, \citenamefont {Ishiwata}, \citenamefont {Matsui},\
  and\ \citenamefont {Tokura}}]{yu2011near}%
  \BibitemOpen
  \bibfield  {author} {\bibinfo {author} {\bibfnamefont {X.}~\bibnamefont
  {Yu}}, \bibinfo {author} {\bibfnamefont {N.}~\bibnamefont {Kanazawa}},
  \bibinfo {author} {\bibfnamefont {Y.}~\bibnamefont {Onose}}, \bibinfo
  {author} {\bibfnamefont {K.}~\bibnamefont {Kimoto}}, \bibinfo {author}
  {\bibfnamefont {W.}~\bibnamefont {Zhang}}, \bibinfo {author} {\bibfnamefont
  {S.}~\bibnamefont {Ishiwata}}, \bibinfo {author} {\bibfnamefont
  {Y.}~\bibnamefont {Matsui}}, \ and\ \bibinfo {author} {\bibfnamefont
  {Y.}~\bibnamefont {Tokura}},\ }\href@noop {} {\bibfield  {journal} {\bibinfo
  {journal} {Nat. Mater.}\ }\textbf {\bibinfo {volume} {10}},\ \bibinfo {pages}
  {106} (\bibinfo {year} {2011})}\BibitemShut {NoStop}%
\bibitem [{\citenamefont {Heinze}\ \emph {et~al.}(2011)\citenamefont {Heinze},
  \citenamefont {Von~Bergmann}, \citenamefont {Menzel}, \citenamefont {Brede},
  \citenamefont {Kubetzka}, \citenamefont {Wiesendanger}, \citenamefont
  {Bihlmayer},\ and\ \citenamefont {Bl{\"u}gel}}]{heinze2011spontaneous}%
  \BibitemOpen
  \bibfield  {author} {\bibinfo {author} {\bibfnamefont {S.}~\bibnamefont
  {Heinze}}, \bibinfo {author} {\bibfnamefont {K.}~\bibnamefont
  {Von~Bergmann}}, \bibinfo {author} {\bibfnamefont {M.}~\bibnamefont
  {Menzel}}, \bibinfo {author} {\bibfnamefont {J.}~\bibnamefont {Brede}},
  \bibinfo {author} {\bibfnamefont {A.}~\bibnamefont {Kubetzka}}, \bibinfo
  {author} {\bibfnamefont {R.}~\bibnamefont {Wiesendanger}}, \bibinfo {author}
  {\bibfnamefont {G.}~\bibnamefont {Bihlmayer}}, \ and\ \bibinfo {author}
  {\bibfnamefont {S.}~\bibnamefont {Bl{\"u}gel}},\ }\href@noop {} {\bibfield
  {journal} {\bibinfo  {journal} {Nat. Phys.}\ }\textbf {\bibinfo {volume}
  {7}},\ \bibinfo {pages} {713} (\bibinfo {year} {2011})}\BibitemShut {NoStop}%
\bibitem [{\citenamefont {Hanneken}\ \emph {et~al.}(2015)\citenamefont
  {Hanneken}, \citenamefont {Otte}, \citenamefont {Kubetzka}, \citenamefont
  {Dup{\'e}}, \citenamefont {Romming}, \citenamefont {von Bergmann},
  \citenamefont {Wiesendanger},\ and\ \citenamefont
  {Heinze}}]{hanneken2015electrical}%
  \BibitemOpen
  \bibfield  {author} {\bibinfo {author} {\bibfnamefont {C.}~\bibnamefont
  {Hanneken}}, \bibinfo {author} {\bibfnamefont {F.}~\bibnamefont {Otte}},
  \bibinfo {author} {\bibfnamefont {A.}~\bibnamefont {Kubetzka}}, \bibinfo
  {author} {\bibfnamefont {B.}~\bibnamefont {Dup{\'e}}}, \bibinfo {author}
  {\bibfnamefont {N.}~\bibnamefont {Romming}}, \bibinfo {author} {\bibfnamefont
  {K.}~\bibnamefont {von Bergmann}}, \bibinfo {author} {\bibfnamefont
  {R.}~\bibnamefont {Wiesendanger}}, \ and\ \bibinfo {author} {\bibfnamefont
  {S.}~\bibnamefont {Heinze}},\ }\href@noop {} {\bibfield  {journal} {\bibinfo
  {journal} {Nat. nanotechnol.}\ }\textbf {\bibinfo {volume} {10}},\ \bibinfo
  {pages} {1039} (\bibinfo {year} {2015})}\BibitemShut {NoStop}%
\bibitem [{\citenamefont {M{\"u}hlbauer}\ \emph {et~al.}(2009)\citenamefont
  {M{\"u}hlbauer}, \citenamefont {Binz}, \citenamefont {Jonietz}, \citenamefont
  {Pfleiderer}, \citenamefont {Rosch}, \citenamefont {Neubauer}, \citenamefont
  {Georgii},\ and\ \citenamefont {B{\"o}ni}}]{muhlbauer2009skyrmion}%
  \BibitemOpen
  \bibfield  {author} {\bibinfo {author} {\bibfnamefont {S.}~\bibnamefont
  {M{\"u}hlbauer}}, \bibinfo {author} {\bibfnamefont {B.}~\bibnamefont {Binz}},
  \bibinfo {author} {\bibfnamefont {F.}~\bibnamefont {Jonietz}}, \bibinfo
  {author} {\bibfnamefont {C.}~\bibnamefont {Pfleiderer}}, \bibinfo {author}
  {\bibfnamefont {A.}~\bibnamefont {Rosch}}, \bibinfo {author} {\bibfnamefont
  {A.}~\bibnamefont {Neubauer}}, \bibinfo {author} {\bibfnamefont
  {R.}~\bibnamefont {Georgii}}, \ and\ \bibinfo {author} {\bibfnamefont
  {P.}~\bibnamefont {B{\"o}ni}},\ }\href@noop {} {\bibfield  {journal}
  {\bibinfo  {journal} {Science}\ }\textbf {\bibinfo {volume} {323}},\ \bibinfo
  {pages} {915} (\bibinfo {year} {2009})}\BibitemShut {NoStop}%
\bibitem [{\citenamefont {Bogdanov}(1995)}]{bogdanov95}%
  \BibitemOpen
  \bibfield  {author} {\bibinfo {author} {\bibfnamefont {A.}~\bibnamefont
  {Bogdanov}},\ }\href@noop {} {\bibfield  {journal} {\bibinfo  {journal} {Sov.
  Phys. JETP Lett.}\ }\textbf {\bibinfo {volume} {62}},\ \bibinfo {pages} {247}
  (\bibinfo {year} {1995})}\BibitemShut {NoStop}%
\bibitem [{\citenamefont {Bogdanov}\ and\ \citenamefont
  {Hubert}(1999)}]{bogdanov1999stability}%
  \BibitemOpen
  \bibfield  {author} {\bibinfo {author} {\bibfnamefont {A.}~\bibnamefont
  {Bogdanov}}\ and\ \bibinfo {author} {\bibfnamefont {A.}~\bibnamefont
  {Hubert}},\ }\href@noop {} {\bibfield  {journal} {\bibinfo  {journal} {J.
  Mag. Mag. Mat.}\ }\textbf {\bibinfo {volume} {195}},\ \bibinfo {pages} {182}
  (\bibinfo {year} {1999})}\BibitemShut {NoStop}%
\bibitem [{\citenamefont {R{\"o}{\ss}ler}\ \emph {et~al.}(2006)\citenamefont
  {R{\"o}{\ss}ler}, \citenamefont {Bogdanov},\ and\ \citenamefont
  {Pfleiderer}}]{rossler2006spontaneous}%
  \BibitemOpen
  \bibfield  {author} {\bibinfo {author} {\bibfnamefont {U.}~\bibnamefont
  {R{\"o}{\ss}ler}}, \bibinfo {author} {\bibfnamefont {A.}~\bibnamefont
  {Bogdanov}}, \ and\ \bibinfo {author} {\bibfnamefont {C.}~\bibnamefont
  {Pfleiderer}},\ }\href@noop {} {\bibfield  {journal} {\bibinfo  {journal}
  {Nature}\ }\textbf {\bibinfo {volume} {442}},\ \bibinfo {pages} {797}
  (\bibinfo {year} {2006})}\BibitemShut {NoStop}%
\bibitem [{\citenamefont {Shibata}\ \emph {et~al.}(2013)\citenamefont
  {Shibata}, \citenamefont {Yu}, \citenamefont {Hara}, \citenamefont
  {Morikawa}, \citenamefont {Kanazawa}, \citenamefont {Kimoto}, \citenamefont
  {Ishiwata}, \citenamefont {Matsui},\ and\ \citenamefont
  {Tokura}}]{shibata2013towards}%
  \BibitemOpen
  \bibfield  {author} {\bibinfo {author} {\bibfnamefont {K.}~\bibnamefont
  {Shibata}}, \bibinfo {author} {\bibfnamefont {X.}~\bibnamefont {Yu}},
  \bibinfo {author} {\bibfnamefont {T.}~\bibnamefont {Hara}}, \bibinfo {author}
  {\bibfnamefont {D.}~\bibnamefont {Morikawa}}, \bibinfo {author}
  {\bibfnamefont {N.}~\bibnamefont {Kanazawa}}, \bibinfo {author}
  {\bibfnamefont {K.}~\bibnamefont {Kimoto}}, \bibinfo {author} {\bibfnamefont
  {S.}~\bibnamefont {Ishiwata}}, \bibinfo {author} {\bibfnamefont
  {Y.}~\bibnamefont {Matsui}}, \ and\ \bibinfo {author} {\bibfnamefont
  {Y.}~\bibnamefont {Tokura}},\ }\href@noop {} {\bibfield  {journal} {\bibinfo
  {journal} {Nat. nanotechnol.}\ }\textbf {\bibinfo {volume} {8}},\ \bibinfo
  {pages} {723} (\bibinfo {year} {2013})}\BibitemShut {NoStop}%
\bibitem [{\citenamefont {Romming}\ \emph {et~al.}(2015)\citenamefont
  {Romming}, \citenamefont {Kubetzka}, \citenamefont {Hanneken}, \citenamefont
  {von Bergmann},\ and\ \citenamefont {Wiesendanger}}]{romming2015field}%
  \BibitemOpen
  \bibfield  {author} {\bibinfo {author} {\bibfnamefont {N.}~\bibnamefont
  {Romming}}, \bibinfo {author} {\bibfnamefont {A.}~\bibnamefont {Kubetzka}},
  \bibinfo {author} {\bibfnamefont {C.}~\bibnamefont {Hanneken}}, \bibinfo
  {author} {\bibfnamefont {K.}~\bibnamefont {von Bergmann}}, \ and\ \bibinfo
  {author} {\bibfnamefont {R.}~\bibnamefont {Wiesendanger}},\ }\href@noop {}
  {\bibfield  {journal} {\bibinfo  {journal} {Phys. Rev. Lett.}\ }\textbf
  {\bibinfo {volume} {114}},\ \bibinfo {pages} {177203} (\bibinfo {year}
  {2015})}\BibitemShut {NoStop}%
\bibitem [{\citenamefont {Hagemeister}\ \emph {et~al.}(2015)\citenamefont
  {Hagemeister}, \citenamefont {Romming}, \citenamefont {von Bergmann},
  \citenamefont {Vedmedenko},\ and\ \citenamefont
  {Wiesendanger}}]{hagemeister2015stability}%
  \BibitemOpen
  \bibfield  {author} {\bibinfo {author} {\bibfnamefont {J.}~\bibnamefont
  {Hagemeister}}, \bibinfo {author} {\bibfnamefont {N.}~\bibnamefont
  {Romming}}, \bibinfo {author} {\bibfnamefont {K.}~\bibnamefont {von
  Bergmann}}, \bibinfo {author} {\bibfnamefont {E.}~\bibnamefont {Vedmedenko}},
  \ and\ \bibinfo {author} {\bibfnamefont {R.}~\bibnamefont {Wiesendanger}},\
  }\href@noop {} {\bibfield  {journal} {\bibinfo  {journal} {Nat. Comm.}\
  }\textbf {\bibinfo {volume} {6}},\ \bibinfo {pages} {8455} (\bibinfo {year}
  {2015})}\BibitemShut {NoStop}%
\bibitem [{\citenamefont {Qi}\ \emph {et~al.}(2006)\citenamefont {Qi},
  \citenamefont {Wu},\ and\ \citenamefont
  {Zhang}}]{QiQuZhang2006TopologicalQuantizationOfTheSpinHallEffectIn2DParamagneticSemiconductors}%
  \BibitemOpen
  \bibfield  {author} {\bibinfo {author} {\bibfnamefont {X.-L.}\ \bibnamefont
  {Qi}}, \bibinfo {author} {\bibfnamefont {Y.-S.}\ \bibnamefont {Wu}}, \ and\
  \bibinfo {author} {\bibfnamefont {S.-C.}\ \bibnamefont {Zhang}},\ }\href
  {\doibase 10.1103/PhysRevB.74.085308} {\bibfield  {journal} {\bibinfo
  {journal} {Phys. Rev. B}\ }\textbf {\bibinfo {volume} {74}},\ \bibinfo
  {pages} {085308} (\bibinfo {year} {2006})}\BibitemShut {NoStop}%
\bibitem [{\citenamefont {Hirsch}(1976)}]{Hirsch1976DifferentialTopology}%
  \BibitemOpen
  \bibfield  {author} {\bibinfo {author} {\bibfnamefont {M.~W.}\ \bibnamefont
  {Hirsch}},\ }\enquote {\bibinfo {title} {Degrees, intersection numbers, and
  the euler characteristic},}\ in\ \href {\doibase 10.1007/978-1-4684-9449-5_6}
  {\emph {\bibinfo {booktitle} {Differential Topology}}}\ (\bibinfo
  {publisher} {Springer New York},\ \bibinfo {address} {New York, NY},\
  \bibinfo {year} {1976})\ pp.\ \bibinfo {pages} {120--141}\BibitemShut
  {NoStop}%
\bibitem [{\citenamefont {Rosch}(2013)}]{rosch2013skyrmions}%
  \BibitemOpen
  \bibfield  {author} {\bibinfo {author} {\bibfnamefont {A.}~\bibnamefont
  {Rosch}},\ }\href@noop {} {\bibfield  {journal} {\bibinfo  {journal} {Nat.
  nanotechnol.}\ }\textbf {\bibinfo {volume} {8}},\ \bibinfo {pages} {160}
  (\bibinfo {year} {2013})}\BibitemShut {NoStop}%
\bibitem [{\citenamefont {Sampaio}\ \emph {et~al.}(2013)\citenamefont
  {Sampaio}, \citenamefont {Cros}, \citenamefont {Rohart}, \citenamefont
  {Thiaville},\ and\ \citenamefont {Fert}}]{sampaio2013nucleation}%
  \BibitemOpen
  \bibfield  {author} {\bibinfo {author} {\bibfnamefont {J.}~\bibnamefont
  {Sampaio}}, \bibinfo {author} {\bibfnamefont {V.}~\bibnamefont {Cros}},
  \bibinfo {author} {\bibfnamefont {S.}~\bibnamefont {Rohart}}, \bibinfo
  {author} {\bibfnamefont {A.}~\bibnamefont {Thiaville}}, \ and\ \bibinfo
  {author} {\bibfnamefont {A.}~\bibnamefont {Fert}},\ }\href@noop {} {\bibfield
   {journal} {\bibinfo  {journal} {Nat. nanotechnol.}\ }\textbf {\bibinfo
  {volume} {8}},\ \bibinfo {pages} {839} (\bibinfo {year} {2013})}\BibitemShut
  {NoStop}%
\bibitem [{\citenamefont {Iwasaki}\ \emph
  {et~al.}(2013{\natexlab{a}})\citenamefont {Iwasaki}, \citenamefont
  {Mochizuki},\ and\ \citenamefont {Nagaosa}}]{iwasaki2013current}%
  \BibitemOpen
  \bibfield  {author} {\bibinfo {author} {\bibfnamefont {J.}~\bibnamefont
  {Iwasaki}}, \bibinfo {author} {\bibfnamefont {M.}~\bibnamefont {Mochizuki}},
  \ and\ \bibinfo {author} {\bibfnamefont {N.}~\bibnamefont {Nagaosa}},\
  }\href@noop {} {\bibfield  {journal} {\bibinfo  {journal} {Nat.
  nanotechnol.}\ }\textbf {\bibinfo {volume} {8}},\ \bibinfo {pages} {742}
  (\bibinfo {year} {2013}{\natexlab{a}})}\BibitemShut {NoStop}%
\bibitem [{\citenamefont {Iwasaki}\ \emph
  {et~al.}(2013{\natexlab{b}})\citenamefont {Iwasaki}, \citenamefont
  {Mochizuki},\ and\ \citenamefont {Nagaosa}}]{iwasaki2013universal}%
  \BibitemOpen
  \bibfield  {author} {\bibinfo {author} {\bibfnamefont {J.}~\bibnamefont
  {Iwasaki}}, \bibinfo {author} {\bibfnamefont {M.}~\bibnamefont {Mochizuki}},
  \ and\ \bibinfo {author} {\bibfnamefont {N.}~\bibnamefont {Nagaosa}},\
  }\href@noop {} {\bibfield  {journal} {\bibinfo  {journal} {Nat. Comm.}\
  }\textbf {\bibinfo {volume} {4}},\ \bibinfo {pages} {1463} (\bibinfo {year}
  {2013}{\natexlab{b}})}\BibitemShut {NoStop}%
\bibitem [{\citenamefont {Jonietz}\ \emph {et~al.}(2010)\citenamefont
  {Jonietz}, \citenamefont {M{\"u}hlbauer}, \citenamefont {Pfleiderer},
  \citenamefont {Neubauer}, \citenamefont {M{\"u}nzer}, \citenamefont {Bauer},
  \citenamefont {Adams}, \citenamefont {Georgii}, \citenamefont {B{\"o}ni},
  \citenamefont {Duine} \emph {et~al.}}]{jonietz2010spin}%
  \BibitemOpen
  \bibfield  {author} {\bibinfo {author} {\bibfnamefont {F.}~\bibnamefont
  {Jonietz}}, \bibinfo {author} {\bibfnamefont {S.}~\bibnamefont
  {M{\"u}hlbauer}}, \bibinfo {author} {\bibfnamefont {C.}~\bibnamefont
  {Pfleiderer}}, \bibinfo {author} {\bibfnamefont {A.}~\bibnamefont
  {Neubauer}}, \bibinfo {author} {\bibfnamefont {W.}~\bibnamefont
  {M{\"u}nzer}}, \bibinfo {author} {\bibfnamefont {A.}~\bibnamefont {Bauer}},
  \bibinfo {author} {\bibfnamefont {T.}~\bibnamefont {Adams}}, \bibinfo
  {author} {\bibfnamefont {R.}~\bibnamefont {Georgii}}, \bibinfo {author}
  {\bibfnamefont {P.}~\bibnamefont {B{\"o}ni}}, \bibinfo {author}
  {\bibfnamefont {R.}~\bibnamefont {Duine}},  \emph {et~al.},\ }\href@noop {}
  {\bibfield  {journal} {\bibinfo  {journal} {Science}\ }\textbf {\bibinfo
  {volume} {330}},\ \bibinfo {pages} {1648} (\bibinfo {year}
  {2010})}\BibitemShut {NoStop}%
\bibitem [{\citenamefont {Yu}\ \emph {et~al.}(2012)\citenamefont {Yu},
  \citenamefont {Kanazawa}, \citenamefont {Zhang}, \citenamefont {Nagai},
  \citenamefont {Hara}, \citenamefont {Kimoto}, \citenamefont {Matsui},
  \citenamefont {Onose},\ and\ \citenamefont {Tokura}}]{yu2012skyrmion}%
  \BibitemOpen
  \bibfield  {author} {\bibinfo {author} {\bibfnamefont {X.}~\bibnamefont
  {Yu}}, \bibinfo {author} {\bibfnamefont {N.}~\bibnamefont {Kanazawa}},
  \bibinfo {author} {\bibfnamefont {W.}~\bibnamefont {Zhang}}, \bibinfo
  {author} {\bibfnamefont {T.}~\bibnamefont {Nagai}}, \bibinfo {author}
  {\bibfnamefont {T.}~\bibnamefont {Hara}}, \bibinfo {author} {\bibfnamefont
  {K.}~\bibnamefont {Kimoto}}, \bibinfo {author} {\bibfnamefont
  {Y.}~\bibnamefont {Matsui}}, \bibinfo {author} {\bibfnamefont
  {Y.}~\bibnamefont {Onose}}, \ and\ \bibinfo {author} {\bibfnamefont
  {Y.}~\bibnamefont {Tokura}},\ }\href@noop {} {\bibfield  {journal} {\bibinfo
  {journal} {Nat. Comm.}\ }\textbf {\bibinfo {volume} {3}},\ \bibinfo {pages}
  {988} (\bibinfo {year} {2012})}\BibitemShut {NoStop}%
\bibitem [{\citenamefont {Krause}\ and\ \citenamefont
  {Wiesendanger}(2016)}]{krause2016spintronics}%
  \BibitemOpen
  \bibfield  {author} {\bibinfo {author} {\bibfnamefont {S.}~\bibnamefont
  {Krause}}\ and\ \bibinfo {author} {\bibfnamefont {R.}~\bibnamefont
  {Wiesendanger}},\ }\href@noop {} {\bibfield  {journal} {\bibinfo  {journal}
  {Nat. Mater.}\ }\textbf {\bibinfo {volume} {15}},\ \bibinfo {pages} {493}
  (\bibinfo {year} {2016})}\BibitemShut {NoStop}%
\bibitem [{\citenamefont {Woo}\ \emph {et~al.}(2016)\citenamefont {Woo},
  \citenamefont {Litzius}, \citenamefont {Kr{\"u}ger}, \citenamefont {Im},
  \citenamefont {Caretta}, \citenamefont {Richter}, \citenamefont {Mann},
  \citenamefont {Krone}, \citenamefont {Reeve}, \citenamefont {Weigand} \emph
  {et~al.}}]{woo2016observation}%
  \BibitemOpen
  \bibfield  {author} {\bibinfo {author} {\bibfnamefont {S.}~\bibnamefont
  {Woo}}, \bibinfo {author} {\bibfnamefont {K.}~\bibnamefont {Litzius}},
  \bibinfo {author} {\bibfnamefont {B.}~\bibnamefont {Kr{\"u}ger}}, \bibinfo
  {author} {\bibfnamefont {M.-Y.}\ \bibnamefont {Im}}, \bibinfo {author}
  {\bibfnamefont {L.}~\bibnamefont {Caretta}}, \bibinfo {author} {\bibfnamefont
  {K.}~\bibnamefont {Richter}}, \bibinfo {author} {\bibfnamefont
  {M.}~\bibnamefont {Mann}}, \bibinfo {author} {\bibfnamefont {A.}~\bibnamefont
  {Krone}}, \bibinfo {author} {\bibfnamefont {R.~M.}\ \bibnamefont {Reeve}},
  \bibinfo {author} {\bibfnamefont {M.}~\bibnamefont {Weigand}},  \emph
  {et~al.},\ }\href@noop {} {\bibfield  {journal} {\bibinfo  {journal} {Nat.
  Mater.}\ ,\ \bibinfo {pages} {501}} (\bibinfo {year} {2016})}\BibitemShut
  {NoStop}%
\bibitem [{\citenamefont {Zhang}\ \emph {et~al.}(2015)\citenamefont {Zhang},
  \citenamefont {Zhao}, \citenamefont {Fangohr}, \citenamefont {Liu},
  \citenamefont {Xia}, \citenamefont {Xia},\ and\ \citenamefont
  {Morvan}}]{zhang2015skyrmion}%
  \BibitemOpen
  \bibfield  {author} {\bibinfo {author} {\bibfnamefont {X.}~\bibnamefont
  {Zhang}}, \bibinfo {author} {\bibfnamefont {G.}~\bibnamefont {Zhao}},
  \bibinfo {author} {\bibfnamefont {H.}~\bibnamefont {Fangohr}}, \bibinfo
  {author} {\bibfnamefont {J.~P.}\ \bibnamefont {Liu}}, \bibinfo {author}
  {\bibfnamefont {W.}~\bibnamefont {Xia}}, \bibinfo {author} {\bibfnamefont
  {J.}~\bibnamefont {Xia}}, \ and\ \bibinfo {author} {\bibfnamefont
  {F.}~\bibnamefont {Morvan}},\ }\href@noop {} {\bibfield  {journal} {\bibinfo
  {journal} {Sci. Rep.}\ }\textbf {\bibinfo {volume} {5}} (\bibinfo {year}
  {2015})}\BibitemShut {NoStop}%
\bibitem [{\citenamefont {Fert}\ \emph {et~al.}(2013)\citenamefont {Fert},
  \citenamefont {Cros},\ and\ \citenamefont {Sampaio}}]{fert2013skyrmions}%
  \BibitemOpen
  \bibfield  {author} {\bibinfo {author} {\bibfnamefont {A.}~\bibnamefont
  {Fert}}, \bibinfo {author} {\bibfnamefont {V.}~\bibnamefont {Cros}}, \ and\
  \bibinfo {author} {\bibfnamefont {J.}~\bibnamefont {Sampaio}},\ }\href@noop
  {} {\bibfield  {journal} {\bibinfo  {journal} {Nat. nanotechnol.}\ }\textbf
  {\bibinfo {volume} {8}},\ \bibinfo {pages} {152} (\bibinfo {year}
  {2013})}\BibitemShut {NoStop}%
\bibitem [{\citenamefont {Nagaosa}\ and\ \citenamefont
  {Tokura}(2013)}]{nagaosa2013topological}%
  \BibitemOpen
  \bibfield  {author} {\bibinfo {author} {\bibfnamefont {N.}~\bibnamefont
  {Nagaosa}}\ and\ \bibinfo {author} {\bibfnamefont {Y.}~\bibnamefont
  {Tokura}},\ }\href@noop {} {\bibfield  {journal} {\bibinfo  {journal} {Nat.
  nanotechnol.}\ }\textbf {\bibinfo {volume} {8}},\ \bibinfo {pages} {899}
  (\bibinfo {year} {2013})}\BibitemShut {NoStop}%
\bibitem [{\citenamefont {Tchoe}\ and\ \citenamefont
  {Han}(2012)}]{tchoe2012skyrmion}%
  \BibitemOpen
  \bibfield  {author} {\bibinfo {author} {\bibfnamefont {Y.}~\bibnamefont
  {Tchoe}}\ and\ \bibinfo {author} {\bibfnamefont {J.~H.}\ \bibnamefont
  {Han}},\ }\href@noop {} {\bibfield  {journal} {\bibinfo  {journal} {Phys.
  Rev. B}\ }\textbf {\bibinfo {volume} {85}},\ \bibinfo {pages} {174416}
  (\bibinfo {year} {2012})}\BibitemShut {NoStop}%
\bibitem [{\citenamefont {Jiang}\ \emph {et~al.}(2015)\citenamefont {Jiang},
  \citenamefont {Upadhyaya}, \citenamefont {Zhang}, \citenamefont {Yu},
  \citenamefont {Jungfleisch}, \citenamefont {Fradin}, \citenamefont {Pearson},
  \citenamefont {Tserkovnyak}, \citenamefont {Wang}, \citenamefont {Heinonen},
  \citenamefont {te~Velthuis},\ and\ \citenamefont {Hoffmann}}]{Jiang283}%
  \BibitemOpen
  \bibfield  {author} {\bibinfo {author} {\bibfnamefont {W.}~\bibnamefont
  {Jiang}}, \bibinfo {author} {\bibfnamefont {P.}~\bibnamefont {Upadhyaya}},
  \bibinfo {author} {\bibfnamefont {W.}~\bibnamefont {Zhang}}, \bibinfo
  {author} {\bibfnamefont {G.}~\bibnamefont {Yu}}, \bibinfo {author}
  {\bibfnamefont {M.~B.}\ \bibnamefont {Jungfleisch}}, \bibinfo {author}
  {\bibfnamefont {F.~Y.}\ \bibnamefont {Fradin}}, \bibinfo {author}
  {\bibfnamefont {J.~E.}\ \bibnamefont {Pearson}}, \bibinfo {author}
  {\bibfnamefont {Y.}~\bibnamefont {Tserkovnyak}}, \bibinfo {author}
  {\bibfnamefont {K.~L.}\ \bibnamefont {Wang}}, \bibinfo {author}
  {\bibfnamefont {O.}~\bibnamefont {Heinonen}}, \bibinfo {author}
  {\bibfnamefont {S.~G.~E.}\ \bibnamefont {te~Velthuis}}, \ and\ \bibinfo
  {author} {\bibfnamefont {A.}~\bibnamefont {Hoffmann}},\ }\href {\doibase
  10.1126/science.aaa1442} {\bibfield  {journal} {\bibinfo  {journal}
  {Science}\ }\textbf {\bibinfo {volume} {349}},\ \bibinfo {pages} {283}
  (\bibinfo {year} {2015})}\BibitemShut {NoStop}%
\bibitem [{\citenamefont {Koshibae}\ and\ \citenamefont
  {Nagaosa}(2016)}]{koshibae2016theory}%
  \BibitemOpen
  \bibfield  {author} {\bibinfo {author} {\bibfnamefont {W.}~\bibnamefont
  {Koshibae}}\ and\ \bibinfo {author} {\bibfnamefont {N.}~\bibnamefont
  {Nagaosa}},\ }\href@noop {} {\bibfield  {journal} {\bibinfo  {journal} {Nat.
  Comm.}\ }\textbf {\bibinfo {volume} {7}} (\bibinfo {year}
  {2016})}\BibitemShut {NoStop}%
\bibitem [{\citenamefont {Romming}\ \emph {et~al.}(2013)\citenamefont
  {Romming}, \citenamefont {Hanneken}, \citenamefont {Menzel}, \citenamefont
  {Bickel}, \citenamefont {Wolter}, \citenamefont {von Bergmann}, \citenamefont
  {Kubetzka},\ and\ \citenamefont {Wiesendanger}}]{romming2013writing}%
  \BibitemOpen
  \bibfield  {author} {\bibinfo {author} {\bibfnamefont {N.}~\bibnamefont
  {Romming}}, \bibinfo {author} {\bibfnamefont {C.}~\bibnamefont {Hanneken}},
  \bibinfo {author} {\bibfnamefont {M.}~\bibnamefont {Menzel}}, \bibinfo
  {author} {\bibfnamefont {J.~E.}\ \bibnamefont {Bickel}}, \bibinfo {author}
  {\bibfnamefont {B.}~\bibnamefont {Wolter}}, \bibinfo {author} {\bibfnamefont
  {K.}~\bibnamefont {von Bergmann}}, \bibinfo {author} {\bibfnamefont
  {A.}~\bibnamefont {Kubetzka}}, \ and\ \bibinfo {author} {\bibfnamefont
  {R.}~\bibnamefont {Wiesendanger}},\ }\href@noop {} {\bibfield  {journal}
  {\bibinfo  {journal} {Science}\ }\textbf {\bibinfo {volume} {341}},\ \bibinfo
  {pages} {636} (\bibinfo {year} {2013})}\BibitemShut {NoStop}%
\bibitem [{\citenamefont {Thiele}(1973)}]{thiele1973steady}%
  \BibitemOpen
  \bibfield  {author} {\bibinfo {author} {\bibfnamefont {A.}~\bibnamefont
  {Thiele}},\ }\href@noop {} {\bibfield  {journal} {\bibinfo  {journal} {Phys.
  Rev. Lett.}\ }\textbf {\bibinfo {volume} {30}},\ \bibinfo {pages} {230}
  (\bibinfo {year} {1973})}\BibitemShut {NoStop}%
\bibitem [{\citenamefont {M{\"u}ller}\ and\ \citenamefont
  {Rosch}(2015)}]{muller2015capturing}%
  \BibitemOpen
  \bibfield  {author} {\bibinfo {author} {\bibfnamefont {J.}~\bibnamefont
  {M{\"u}ller}}\ and\ \bibinfo {author} {\bibfnamefont {A.}~\bibnamefont
  {Rosch}},\ }\href@noop {} {\bibfield  {journal} {\bibinfo  {journal} {Phys.
  Rev. B}\ }\textbf {\bibinfo {volume} {91}},\ \bibinfo {pages} {054410}
  (\bibinfo {year} {2015})}\BibitemShut {NoStop}%
\bibitem [{\citenamefont {Everschor-Sitte}\ \emph {et~al.}(2016)\citenamefont
  {Everschor-Sitte}, \citenamefont {Sitte}, \citenamefont {Valet},
  \citenamefont {Sinova},\ and\ \citenamefont
  {Abanov}}]{everschor2016skyrmion}%
  \BibitemOpen
  \bibfield  {author} {\bibinfo {author} {\bibfnamefont {K.}~\bibnamefont
  {Everschor-Sitte}}, \bibinfo {author} {\bibfnamefont {M.}~\bibnamefont
  {Sitte}}, \bibinfo {author} {\bibfnamefont {T.}~\bibnamefont {Valet}},
  \bibinfo {author} {\bibfnamefont {J.}~\bibnamefont {Sinova}}, \ and\ \bibinfo
  {author} {\bibfnamefont {A.}~\bibnamefont {Abanov}},\ }\href@noop {}
  {\bibfield  {journal} {\bibinfo  {journal} {arXiv:1610.08313}\ } (\bibinfo
  {year} {2016})}\BibitemShut {NoStop}%
\bibitem [{\citenamefont {Yu}\ \emph {et~al.}(2017)\citenamefont {Yu},
  \citenamefont {Morikawa}, \citenamefont {Tokunaga}, \citenamefont {Kubota},
  \citenamefont {Kurumaji}, \citenamefont {Oike}, \citenamefont {Nakamura},
  \citenamefont {Kagawa}, \citenamefont {Taguchi}, \citenamefont {Arima},
  \citenamefont {Kawasaki},\ and\ \citenamefont {Tokura}}]{yu2017}%
  \BibitemOpen
  \bibfield  {author} {\bibinfo {author} {\bibfnamefont {X.}~\bibnamefont
  {Yu}}, \bibinfo {author} {\bibfnamefont {D.}~\bibnamefont {Morikawa}},
  \bibinfo {author} {\bibfnamefont {Y.}~\bibnamefont {Tokunaga}}, \bibinfo
  {author} {\bibfnamefont {M.}~\bibnamefont {Kubota}}, \bibinfo {author}
  {\bibfnamefont {T.}~\bibnamefont {Kurumaji}}, \bibinfo {author}
  {\bibfnamefont {H.}~\bibnamefont {Oike}}, \bibinfo {author} {\bibfnamefont
  {M.}~\bibnamefont {Nakamura}}, \bibinfo {author} {\bibfnamefont
  {F.}~\bibnamefont {Kagawa}}, \bibinfo {author} {\bibfnamefont
  {Y.}~\bibnamefont {Taguchi}}, \bibinfo {author} {\bibfnamefont {T.-h.}\
  \bibnamefont {Arima}}, \bibinfo {author} {\bibfnamefont {M.}~\bibnamefont
  {Kawasaki}}, \ and\ \bibinfo {author} {\bibfnamefont {Y.}~\bibnamefont
  {Tokura}},\ }\href {\doibase 10.1002/adma.201606178} {\bibfield  {journal}
  {\bibinfo  {journal} {Adv. Mat.}\ ,\ \bibinfo {pages} {1606178}} (\bibinfo
  {year} {2017})},\ \bibinfo {note} {1606178}\BibitemShut {NoStop}%
\bibitem [{\citenamefont {Tatara}\ \emph {et~al.}(2008)\citenamefont {Tatara},
  \citenamefont {Kohno},\ and\ \citenamefont
  {Shibata}}]{tatara2008microscopic}%
  \BibitemOpen
  \bibfield  {author} {\bibinfo {author} {\bibfnamefont {G.}~\bibnamefont
  {Tatara}}, \bibinfo {author} {\bibfnamefont {H.}~\bibnamefont {Kohno}}, \
  and\ \bibinfo {author} {\bibfnamefont {J.}~\bibnamefont {Shibata}},\
  }\href@noop {} {\bibfield  {journal} {\bibinfo  {journal} {Phys. Rep.}\
  }\textbf {\bibinfo {volume} {468}},\ \bibinfo {pages} {213} (\bibinfo {year}
  {2008})}\BibitemShut {NoStop}%
\bibitem [{\citenamefont {Li}\ and\ \citenamefont
  {Zhang}(2004)}]{li2004domain}%
  \BibitemOpen
  \bibfield  {author} {\bibinfo {author} {\bibfnamefont {Z.}~\bibnamefont
  {Li}}\ and\ \bibinfo {author} {\bibfnamefont {S.}~\bibnamefont {Zhang}},\
  }\href@noop {} {\bibfield  {journal} {\bibinfo  {journal} {Phys. Rev. Lett.}\
  }\textbf {\bibinfo {volume} {92}},\ \bibinfo {pages} {207203} (\bibinfo
  {year} {2004})}\BibitemShut {NoStop}%
\bibitem [{\citenamefont {Bazaliy}\ \emph {et~al.}(1998)\citenamefont
  {Bazaliy}, \citenamefont {Jones},\ and\ \citenamefont
  {Zhang}}]{bazaliy1998modification}%
  \BibitemOpen
  \bibfield  {author} {\bibinfo {author} {\bibfnamefont {Y.~B.}\ \bibnamefont
  {Bazaliy}}, \bibinfo {author} {\bibfnamefont {B.}~\bibnamefont {Jones}}, \
  and\ \bibinfo {author} {\bibfnamefont {S.-C.}\ \bibnamefont {Zhang}},\
  }\href@noop {} {\bibfield  {journal} {\bibinfo  {journal} {Phys. Rev. B}\
  }\textbf {\bibinfo {volume} {57}},\ \bibinfo {pages} {R3213} (\bibinfo {year}
  {1998})}\BibitemShut {NoStop}%
\bibitem [{\citenamefont {Lakshmanan}(2011)}]{lakshmanan2011fascinating}%
  \BibitemOpen
  \bibfield  {author} {\bibinfo {author} {\bibfnamefont {M.}~\bibnamefont
  {Lakshmanan}},\ }\href@noop {} {\bibfield  {journal} {\bibinfo  {journal}
  {Philosophical Transactions of the Royal Society of London A: Mathematical,
  Physical and Engineering Sciences}\ }\textbf {\bibinfo {volume} {369}},\
  \bibinfo {pages} {1280} (\bibinfo {year} {2011})}\BibitemShut {NoStop}%
\bibitem [{\citenamefont {Zhang}\ and\ \citenamefont
  {Zhang}(2009)}]{zhang2009generalization}%
  \BibitemOpen
  \bibfield  {author} {\bibinfo {author} {\bibfnamefont {S.}~\bibnamefont
  {Zhang}}\ and\ \bibinfo {author} {\bibfnamefont {S.~S.-L.}\ \bibnamefont
  {Zhang}},\ }\href@noop {} {\bibfield  {journal} {\bibinfo  {journal} {Phys.
  Rev. Lett.}\ }\textbf {\bibinfo {volume} {102}},\ \bibinfo {pages} {086601}
  (\bibinfo {year} {2009})}\BibitemShut {NoStop}%
\bibitem [{\citenamefont {Heo}\ \emph {et~al.}(2016)\citenamefont {Heo},
  \citenamefont {Kiselev}, \citenamefont {Nandy}, \citenamefont {Bl{\"u}gel},\
  and\ \citenamefont {Rasing}}]{heo2016switching}%
  \BibitemOpen
  \bibfield  {author} {\bibinfo {author} {\bibfnamefont {C.}~\bibnamefont
  {Heo}}, \bibinfo {author} {\bibfnamefont {N.~S.}\ \bibnamefont {Kiselev}},
  \bibinfo {author} {\bibfnamefont {A.~K.}\ \bibnamefont {Nandy}}, \bibinfo
  {author} {\bibfnamefont {S.}~\bibnamefont {Bl{\"u}gel}}, \ and\ \bibinfo
  {author} {\bibfnamefont {T.}~\bibnamefont {Rasing}},\ }\href@noop {}
  {\bibfield  {journal} {\bibinfo  {journal} {Sci. Rep.}\ }\textbf {\bibinfo
  {volume} {6}} (\bibinfo {year} {2016})}\BibitemShut {NoStop}%
\bibitem [{\citenamefont {Bogdanov}\ and\ \citenamefont
  {Yablonskii}(1989)}]{bogdanov1989thermodynamically}%
  \BibitemOpen
  \bibfield  {author} {\bibinfo {author} {\bibfnamefont {A.}~\bibnamefont
  {Bogdanov}}\ and\ \bibinfo {author} {\bibfnamefont {D.}~\bibnamefont
  {Yablonskii}},\ }\href@noop {} {\bibfield  {journal} {\bibinfo  {journal}
  {Zh. Eksp. Teor. Fiz}\ }\textbf {\bibinfo {volume} {95}},\ \bibinfo {pages}
  {182} (\bibinfo {year} {1989})}\BibitemShut {NoStop}%
\bibitem [{\citenamefont {Garst}(2016)}]{Garst2016}%
  \BibitemOpen
  \bibfield  {author} {\bibinfo {author} {\bibfnamefont {M.}~\bibnamefont
  {Garst}},\ }\enquote {\bibinfo {title} {Topological skyrmion dynamics in
  chiral magnets},}\ in\ \href {\doibase 10.1007/978-3-319-25301-5_2} {\emph
  {\bibinfo {booktitle} {Topological Structures in Ferroic Materials: Domain
  Walls, Vortices and Skyrmions}}},\ \bibinfo {editor} {edited by\ \bibinfo
  {editor} {\bibfnamefont {J.}~\bibnamefont {Seidel}}}\ (\bibinfo  {publisher}
  {Springer International Publishing},\ \bibinfo {address} {Cham},\ \bibinfo
  {year} {2016})\ pp.\ \bibinfo {pages} {29--53}\BibitemShut {NoStop}%
\bibitem [{\citenamefont {Frederico}\ and\ \citenamefont
  {Torres}(2007)}]{FREDERICO2007834}%
  \BibitemOpen
  \bibfield  {author} {\bibinfo {author} {\bibfnamefont {G.~S.}\ \bibnamefont
  {Frederico}}\ and\ \bibinfo {author} {\bibfnamefont {D.~F.}\ \bibnamefont
  {Torres}},\ }\href {\doibase http://dx.doi.org/10.1016/j.jmaa.2007.01.013}
  {\bibfield  {journal} {\bibinfo  {journal} {J. Math. Anal. Appl.}\ }\textbf
  {\bibinfo {volume} {334}},\ \bibinfo {pages} {834 } (\bibinfo {year}
  {2007})}\BibitemShut {NoStop}%
\bibitem [{\citenamefont {Stone}(1996)}]{stone1996}%
  \BibitemOpen
  \bibfield  {author} {\bibinfo {author} {\bibfnamefont {M.}~\bibnamefont
  {Stone}},\ }\href {\doibase 10.1103/PhysRevB.53.16573} {\bibfield  {journal}
  {\bibinfo  {journal} {Phys. Rev. B}\ }\textbf {\bibinfo {volume} {53}},\
  \bibinfo {pages} {16573} (\bibinfo {year} {1996})}\BibitemShut {NoStop}%
\bibitem [{\citenamefont {Nakatani}\ \emph {et~al.}(2008)\citenamefont
  {Nakatani}, \citenamefont {Shibata}, \citenamefont {Tatara}, \citenamefont
  {Kohno}, \citenamefont {Thiaville},\ and\ \citenamefont
  {Miltat}}]{Nakatani2008}%
  \BibitemOpen
  \bibfield  {author} {\bibinfo {author} {\bibfnamefont {Y.}~\bibnamefont
  {Nakatani}}, \bibinfo {author} {\bibfnamefont {J.}~\bibnamefont {Shibata}},
  \bibinfo {author} {\bibfnamefont {G.}~\bibnamefont {Tatara}}, \bibinfo
  {author} {\bibfnamefont {H.}~\bibnamefont {Kohno}}, \bibinfo {author}
  {\bibfnamefont {A.}~\bibnamefont {Thiaville}}, \ and\ \bibinfo {author}
  {\bibfnamefont {J.}~\bibnamefont {Miltat}},\ }\href {\doibase
  10.1103/PhysRevB.77.014439} {\bibfield  {journal} {\bibinfo  {journal} {Phys.
  Rev. B}\ }\textbf {\bibinfo {volume} {77}},\ \bibinfo {pages} {014439}
  (\bibinfo {year} {2008})}\BibitemShut {NoStop}%
\bibitem [{\citenamefont {Zhang}\ \emph {et~al.}(2016)\citenamefont {Zhang},
  \citenamefont {Yu}, \citenamefont {Zhang}, \citenamefont {Wang},
  \citenamefont {Benjamin~Jungfleisch}, \citenamefont {Pearson}, \citenamefont
  {Cheng}, \citenamefont {Heinonen}, \citenamefont {Wang}, \citenamefont
  {Zhou}, \citenamefont {Hoffmann},\ and\ \citenamefont
  {te~Velthuis}}]{jiang2016direct}%
  \BibitemOpen
  \bibfield  {author} {\bibinfo {author} {\bibfnamefont {X.}~\bibnamefont
  {Zhang}}, \bibinfo {author} {\bibfnamefont {G.}~\bibnamefont {Yu}}, \bibinfo
  {author} {\bibfnamefont {W.}~\bibnamefont {Zhang}}, \bibinfo {author}
  {\bibfnamefont {X.}~\bibnamefont {Wang}}, \bibinfo {author} {\bibfnamefont
  {M.}~\bibnamefont {Benjamin~Jungfleisch}}, \bibinfo {author} {\bibfnamefont
  {J.~E.}\ \bibnamefont {Pearson}}, \bibinfo {author} {\bibfnamefont
  {X.}~\bibnamefont {Cheng}}, \bibinfo {author} {\bibfnamefont
  {O.}~\bibnamefont {Heinonen}}, \bibinfo {author} {\bibfnamefont {K.~L.}\
  \bibnamefont {Wang}}, \bibinfo {author} {\bibfnamefont {Y.}~\bibnamefont
  {Zhou}}, \bibinfo {author} {\bibfnamefont {A.}~\bibnamefont {Hoffmann}}, \
  and\ \bibinfo {author} {\bibfnamefont {S.~G.~E.}\ \bibnamefont
  {te~Velthuis}},\ }\href@noop {} {\bibfield  {journal} {\bibinfo  {journal}
  {Nat. Phys.}\ }\textbf {\bibinfo {volume} {13}},\ \bibinfo {pages}
  {162–169} (\bibinfo {year} {2016})}\BibitemShut {NoStop}%
\bibitem [{\citenamefont {Litzius}\ \emph {et~al.}(2016)\citenamefont
  {Litzius}, \citenamefont {Lemesh}, \citenamefont {Kr{\"u}ger}, \citenamefont
  {Bassirian}, \citenamefont {Caretta}, \citenamefont {Richter}, \citenamefont
  {B{\"u}ttner}, \citenamefont {Sato}, \citenamefont {Tretiakov}, \citenamefont
  {F{\"o}rster}, \citenamefont {Reeve}, \citenamefont {Weigand}, \citenamefont
  {Bykova}, \citenamefont {Stoll}, \citenamefont {Schutz}, \citenamefont
  {Beach},\ and\ \citenamefont {Kl{\"a}ui}}]{litzius2016skyrmion}%
  \BibitemOpen
  \bibfield  {author} {\bibinfo {author} {\bibfnamefont {K.}~\bibnamefont
  {Litzius}}, \bibinfo {author} {\bibfnamefont {I.}~\bibnamefont {Lemesh}},
  \bibinfo {author} {\bibfnamefont {B.}~\bibnamefont {Kr{\"u}ger}}, \bibinfo
  {author} {\bibfnamefont {P.}~\bibnamefont {Bassirian}}, \bibinfo {author}
  {\bibfnamefont {L.}~\bibnamefont {Caretta}}, \bibinfo {author} {\bibfnamefont
  {K.}~\bibnamefont {Richter}}, \bibinfo {author} {\bibfnamefont
  {F.}~\bibnamefont {B{\"u}ttner}}, \bibinfo {author} {\bibfnamefont
  {K.}~\bibnamefont {Sato}}, \bibinfo {author} {\bibfnamefont {O.~A.}\
  \bibnamefont {Tretiakov}}, \bibinfo {author} {\bibfnamefont {J.}~\bibnamefont
  {F{\"o}rster}}, \bibinfo {author} {\bibfnamefont {R.~M.}\ \bibnamefont
  {Reeve}}, \bibinfo {author} {\bibfnamefont {M.}~\bibnamefont {Weigand}},
  \bibinfo {author} {\bibfnamefont {I.}~\bibnamefont {Bykova}}, \bibinfo
  {author} {\bibfnamefont {H.}~\bibnamefont {Stoll}}, \bibinfo {author}
  {\bibfnamefont {G.}~\bibnamefont {Schutz}}, \bibinfo {author} {\bibfnamefont
  {G.~S.~D.}\ \bibnamefont {Beach}}, \ and\ \bibinfo {author} {\bibfnamefont
  {M.}~\bibnamefont {Kl{\"a}ui}},\ }\href@noop {} {\bibfield  {journal}
  {\bibinfo  {journal} {Nat. Phys.}\ }\textbf {\bibinfo {volume} {13}},\
  \bibinfo {pages} {170–175} (\bibinfo {year} {2016})}\BibitemShut {NoStop}%
\bibitem [{\citenamefont {Thiaville}\ \emph {et~al.}(2012)\citenamefont
  {Thiaville}, \citenamefont {Rohart}, \citenamefont {Ju{\'e}}, \citenamefont
  {Cros},\ and\ \citenamefont {Fert}}]{thiaville2012dynamics}%
  \BibitemOpen
  \bibfield  {author} {\bibinfo {author} {\bibfnamefont {A.}~\bibnamefont
  {Thiaville}}, \bibinfo {author} {\bibfnamefont {S.}~\bibnamefont {Rohart}},
  \bibinfo {author} {\bibfnamefont {{\'E}.}~\bibnamefont {Ju{\'e}}}, \bibinfo
  {author} {\bibfnamefont {V.}~\bibnamefont {Cros}}, \ and\ \bibinfo {author}
  {\bibfnamefont {A.}~\bibnamefont {Fert}},\ }\href@noop {} {\bibfield
  {journal} {\bibinfo  {journal} {Europhys. Lett.}\ }\textbf {\bibinfo {volume}
  {100}},\ \bibinfo {pages} {57002} (\bibinfo {year} {2012})}\BibitemShut
  {NoStop}%
\bibitem [{\citenamefont {Benitez}\ \emph {et~al.}(2015)\citenamefont
  {Benitez}, \citenamefont {Hrabec}, \citenamefont {Mihai}, \citenamefont
  {Moore}, \citenamefont {Burnell}, \citenamefont {McGrouther}, \citenamefont
  {Marrows},\ and\ \citenamefont {McVitie}}]{benitez2015magnetic}%
  \BibitemOpen
  \bibfield  {author} {\bibinfo {author} {\bibfnamefont {M.}~\bibnamefont
  {Benitez}}, \bibinfo {author} {\bibfnamefont {A.}~\bibnamefont {Hrabec}},
  \bibinfo {author} {\bibfnamefont {A.}~\bibnamefont {Mihai}}, \bibinfo
  {author} {\bibfnamefont {T.}~\bibnamefont {Moore}}, \bibinfo {author}
  {\bibfnamefont {G.}~\bibnamefont {Burnell}}, \bibinfo {author} {\bibfnamefont
  {D.}~\bibnamefont {McGrouther}}, \bibinfo {author} {\bibfnamefont
  {C.}~\bibnamefont {Marrows}}, \ and\ \bibinfo {author} {\bibfnamefont
  {S.}~\bibnamefont {McVitie}},\ }\href@noop {} {\bibfield  {journal} {\bibinfo
   {journal} {Nat. Comm.}\ }\textbf {\bibinfo {volume} {6}} (\bibinfo {year}
  {2015})}\BibitemShut {NoStop}%
\bibitem [{SM()}]{SM}%
  \BibitemOpen
  \href@noop {} {}\bibinfo {note} {See Supplemental Material at the ancillary file section of 
arXiv (``other formats'') for movies of skyrmion-pair evolution.}\BibitemShut {Stop}%
\end{thebibliography}
\end{document}